\documentclass[12pt,preprint]{aastex}

\def\VEC#1{\mbox{\boldmath $#1$}}

\slugcomment{Not to appear in Nonlearned J., 45.}

\shorttitle{Generalized RMHD Equations for Pair and Electron-Ion Plasmas}
\shortauthors{Koide}

\begin{document}

\title{Generalized Relativistic Magnetohydrodynamic Equations \\
     for Pair and Electron-Ion Plasmas}

\author{Shinji Koide}
\affil{Department of Physics, Kumamoto University,
    2-39-1, Kurokami, Kumamoto, 860-8555, Japan}
\email{koidesin@sci.kumamoto-u.ac.jp}

\begin{abstract}
We derived one-fluid equations based on a relativistic two-fluid approximation
of e$^\pm$ pair plasma and electron-ion plasma to reveal 
the specific relativistic nature of their behavior. 
Assuming simple condition on the relativistic one-fluid
equations, we propose generalized relativistic magnetohydrodynamic (RMHD) 
equations which satisfy causality.
We show the linear analyses of these equations regarding various plasma 
waves to show the validity of the generalized RMHD equations derived here and 
to reveal the distinct properties of the pair plasma and electron-ion plasma. 
The distinct properties relate to (i) the inertia effect of 
electric charge,
(ii) the momentum of electric current, (iii) the relativistic Hall effect, 
(iv) the thermal electromotive force, and (v) the thermalized energy exchange 
between the two fluids. Using the generalized RMHD equations, we also clarify 
the condition that we can
use standard RMHD equations and that we need the distinct RMHD equations
of pair and electron-ion plasmas. The standard RMHD is available only
when the relative velocity of the two fluids is nonrelativistic,
a difference of the enthalpy densities of the two fluids is much
smaller than the total enthalpy density, and the above distinct
properties of the pair/electron-ion plasma are negligible.
We discuss a general relativistic version of the equations applicable to 
the pair and electron-ion plasmas in black hole magnetospheres. 
We find the effective resistivity due to shear of frame dragging
around a rotating black hole.
\end{abstract}


\section{Introduction}

The most powerful engines in Universe are located in centers of 
active galactic nuclei (AGNs), quasars (QSOs), 
micro quasars ($\mu$QSOs), and gamma-ray bursts (GRBs).
Observations have shown all of them eject relativistic jets
\cite{pearson87,biretta99,junor99,mirabel94,mirabel98,kulkarni99}.
In spite of the drastic difference of their characteristic scales and powers,
it is believed that the activities of these objects, such as the relativistic
jet ejection, are supported commonly by
the drastic phenomena of accretion disks around black holes 
\cite{mirabel98}.
However, the distinct mechanism of the activity 
is not confirmed yet.
Models with interaction between the plasma and magnetic field
in the very strong gravity of the black hole are thought to be most promising.
The region where the plasma and magnetic field interact each other
near a black hole is called a black hole magnetosphere.
A black hole magnetosphere consists of an accretion disk and 
a corona around a black hole and it is thought to be composed of various 
kinds of plasmas. In a case of an AGN, it is suggested 
that the disk is made of electron-ion plasma \cite{ford94}, the core of the 
relativistic jet is mainly of electron-positron (pair) 
plasma \citep{wardle98},
and the corona is by both of them, while the actual components of such plasmas 
have not been confirmed observationally yet.
It is a natural idea that
a corona near a black hole consists mainly of pair plasma,
and near its disk, mainly of electron-ion plasmas.
Such a difference of the kinds of plasmas may influence
the dynamics of a black hole magnetosphere, but none of the works has been
done to investigate it.

Black hole magnetosphere has been studied on the basis of
the non-relativistic magnetohydrodynamics (MHD) with 
pseudo-Newtonian potential \cite{paczynski80}
and ideal general relativistic MHD (GRMHD) 
\cite[e.g.,][]{takahashi90,tomimatsu94}.
The relativistic MHD (RMHD) or GRMHD is a one-fluid
approximation of the plasma and is based on 
the relativistic conservation laws of particle number, momentum, and energy, 
Maxwell equations, and the simple Ohm's law in the plasma rest 
frame. The simple Ohm's law means that the electric
field $\VEC{E}'$ is proportional to the current density $\VEC{J}'$ 
in the plasma rest frame: $\VEC{E}' = \eta \VEC{J}'$
where $\eta$ is the electric resistivity.
We call such RMHD (GRMHD) equations the ``standard" RMHD (GRMHD)
equations in this paper. 
When we set $\eta$ zero, the standard RMHD (GRMHD) becomes 
the ``ideal" RMHD (GRMHD).
Recently, numerical simulations of the ideal GRMHD have
become prevailing and yielded interesting, important results
with respect to relativistic outflow/jet formation \cite{koide04,mckinney06} and
extraction of black hole rotational energy by magnetic
field \cite{koide02,koide03,komissarov05}.
However, since they used the standard GRMHD and thus the one-fluid
approximation,
they ignored the unique properties of the pair/electron-ion plasma.
To begin with, we have to check the validity of the 
standard RMHD and clarify its application conditions
for plasmas around black holes.

To clarify these points, we reconsider the RMHD equations.
Such a task was first performed by Ardavan (1976) using the 
Vlasov--Boltzmann equation for a pulsar magnetosphere. It yielded a 
relativistic version of the generalized Ohm's law and a new condition 
for the validity of the MHD approximation for a pulsar magnetosphere 
(where Lorentz factor of plasma is much larger than unity). A more generalized 
treatment, which included annihilation of electrons and positrons, 
radiation, Compton scattering, and pair photoproduction was formulated 
by Blackman \& Field (1993) and Gedalin (1996). Reconsideration of ideal MHD 
in a neutral cold plasma based on two-fluid approximation was presented 
by Melatos \& Melrose (1996), who investigated the conditions under which the MHD 
approximation breaks down. To investigate black hole magnetospheres,
Khanna (1998) formulated a general relativistic version of the 
two-fluid approximation in the Kerr metric. A more generalized 
version in a time-varying space-time was derived by Meier (2004) from 
the general relativistic Vlasov--Boltzmann equation.

To investigate the distinct properties of the pair and electron-ion plasmas and 
the applicability of the standard RMHD, 
we derived one-fluid equations
using a relativistic two-fluid approximation of a plasma consisting
of positively charged particles, each has charge $e$ and mass $m_+$,
and the other fluid of negative particles, each with charge $-e$ and mass $m_-$.
Throughout this paper, we assume the charge of ions is $e$, while
generalization for plasmas with arbitrary charge of ion is not difficult.
In an electron-ion plasma case, $m_+$ is much larger than
$m_-$, while in a pair plasma case, they are equal.
With respect to the one-fluid equations with finite resistivity
larger than a certain value, 
causality seems to be broken mathematically \cite{koide08b}.
We consider the dispersion relation of the electromagnetic waves
and show the mathematical condition where the equations satisfy causality.

On the basis of the above results, we propose the ``generalized RMHD equations"
for the plasmas, which are introduced in this field for the first time 
by the present work. These equations can reveal the
distinct properties of the electron-ion and pair plasmas. 
The differences between the generalized RMHD equations and the standard
RMHD equations concern (i) the inertia effect of charge, (ii) the momentum of
current, (iii) the relativistic Hall effect, 
(iv) the thermal electromotive force, and
(v) the thermalized energy exchange rate. Mainly using linear
analyses, we investigate the influence of such distinct properties to plasma dynamics
which cannot be clarified from the standard RMHD equations. 
The conditions that the distinct properties of the electron-ion 
and pair plasmas are negligible
and the premise conditions
of the generalized RMHD clarify the application condition of the standard 
RMHD equations. 

In section \ref{sec2}, we derive one-fluid equations from the relativistic 
two-fluid
equations. In section \ref{sec3}, dispersion relation of electromagnetic 
wave is shown on the basis of the one-fluid equations and we confirm 
the mathematical condition
of causality of the one-fluid equations.
In section \ref{sec4}, we propose a set of generalized RMHD equations with
any mass ratio $m_+/m_-$ from the one-fluid equations. 
To show the reasonable property of the generalized RMHD equations,
in section \ref{sec5}, we investigate the linear
modes of various plasma waves derived by the generalized RMHD equations 
and show the unique nature of the pair and electron-ion plasmas.
Section \ref{sec6} shows the unique properties of the electron-ion and pair
plasmas and estimations of them.
In section \ref{sec7}, we discuss the break-down condition of 
the generalized RMHD equations and the non-linear distinct nature of the
electron-ion and pair plasmas in the jet forming region around
the black hole. 
In section \ref{sec8}, we suggest the distinct general relativistic effects 
shown by the generalized GRMHD equations.
Here, we indicate that shear of frame dragging induces 
effective resistivity near a rotating black hole.
In section \ref{sec9}, the results are summarized.

\section{Two-fluid equations and one-fluid equations
\label{sec2}}

We employ relativistic two-fluid equations of plasma consisting of
two fluids. One fluid is composed of positively charged particles,
 and the other is composed of 
the negatively charged particles.
We derive a set of one-fluid equations equivalent to the two-fluid
equations in the Minkowski space-time $(x^0,x^1,x^2,x^3)=(t,x,y,z)$
where the line element is given by 
$ds^2=-(dx^0)^2+(dx^1)^2+(dx^2)^2+(dx^3)^2=\eta_{\mu\nu}dx^\mu dx^\nu$.
Throughout this paper, we use units in which the speed of light,
the dielectric constant, and the magnetic permeability in vacuum
all are unity: $c=1$, $\epsilon_0=1$, $\mu_0=1$.
The relativistic equations of the two fluids and the Maxwell equations
are 
\begin{eqnarray} 
\partial_\nu (n_\pm U_\pm^\nu) &=& 0 , \label{4formnum} \\ 
\partial_\nu (h_\pm U_\pm^\mu U_\pm^\nu)  &=& 
-\partial^\mu p_\pm \pm e n_\pm \eta^{\mu\sigma} U_\pm^\nu F_{\sigma\nu} 
\pm R^\mu , \label{4formmom} \\ 
\partial_\nu \hspace{0.3em} ^*F^{\mu\nu} &=& 0 , \label{4formfar} \\ 
\partial_\nu F^{\mu\nu} & = & J^\mu , \label{4formamp} 
\end{eqnarray} 
where variables with subscripts, plus (+) and minus (--), 
are those of the fluid of positively charged particles and of the fluid 
of negative particles, respectively, 
$n_\pm$ is the proper particle number density, $U_\pm^\mu$ is the 
four-velocity, $p_\pm$ is 
the proper pressure, $h_\pm$ is the relativistic enthalpy density, 
$F_{\mu\nu}$ is the electromagnetic field tensor, 
$\hspace{0.3em} ^*F^{\mu\nu}$ is the dual tensor density of $F_{\mu\nu}$, 
$R^\mu$ is the frictional four-force density between the two fluids, 
and $J^\mu$ is the four-current density. 
We will often write a set of the spacial components of the four-vector 
using a bold italic font, e.g., $\VEC{U}_\pm = (U^1_\pm,U^2_\pm,U^3_\pm)$, 
$\VEC{J} = (J^1,J^2,J^3)$, $\VEC{R} = (R^1, R^2, R^3)$. 
We further define the Lorentz factor $\gamma_\pm = U^0_\pm$, the three-velocity 
$V^i_\pm=U^i_\pm/\gamma_\pm$, the electric field $E_i=F^{0i}$, the magnetic flux 
density $B_i=\sum_{jk} \epsilon_{ijk} F^{jk}/2$ 
($\epsilon_{ijk}$ is the Levi--Civita tensor), and the electric charge density 
$\rho_{\rm e} = J^0$. 
Here, the alphabetic index ($i,j,k$) runs from 1 to 3.
Throughout this paper, we assume that the two fluids are heated 
only by Ohmic heating and disregard nuclear reactions and pair creation 
and annihilation. We also ignore radiation and quantum effects.

To derive one-fluid equations of the plasma, we define the average and 
difference variables as follows: 
\begin{eqnarray} 
\rho &=& m_+ n_+ + m_- n_- , \label{averho}\\
n &=& \frac{\rho}{m} ,\\
U^\mu &=& \frac{1}{\rho} ( m_+ n_+ U_+^\mu + m_- n_- U_-^\mu ) , 
\label{ave4vel} \\ 
J^\mu &=& e(n_+ U_+^\mu - n_- U_-^\mu) , 
\label{ave4cur}
\end{eqnarray}
where $m=m_+ + m_-$.
Then we can write
\begin{equation}
n_\pm U_\pm ^\mu = \frac{1}{m} \left  ( 
\rho U^\mu \pm \frac{m_\mp}{e} J^\mu \right )  .
\end{equation}
We also define the average and difference variables with respect to the
enthalpy density as
\begin{eqnarray} 
h^\dagger &=& n^2 \left ( \frac{h_+}{n_+^2} + \frac{h_-}{n_-^2} \right ), 
\label{aveenth1} \\ 
\Delta h^\dagger &=&n^2 \left ( 
\frac{h_+}{n_+^2} \frac{2m_-}{m} - \frac{h_-}{n_-^2} \frac{2m_+}{m} \right ), \\ 
h^\ddagger &=&n^2 \left [ 
\frac{h_+}{n_+^2} \left ( \frac{2m_-}{m} \right )^2 
+ \frac{h_-}{n_-^2} \left ( \frac{2m_+}{m} \right )^2 \right ], \\ 
\Delta h^\ddagger &=&n^2 \left ( 
\frac{h_+}{n_+^2} \frac{2m_+}{m} - \frac{h_-}{n_-^2} \frac{2m_-}{m} \right ), \\ 
\Delta h^\sharp &=&n^2 \left [ 
\frac{h_+}{n_+^2} \left ( \frac{2m_-}{m} \right )^3 
- \frac{h_-}{n_-^2} \left ( \frac{2m_+}{m} \right )^3 \right ] . 
\end{eqnarray}
We find the following relations between the variables with respect to the
enthalpy density,
\begin{eqnarray}
h^\ddagger &=& 4 \mu h^\dagger - \Delta \mu \Delta h^\dagger , \\
\Delta h^\ddagger &=& 2 \Delta \mu h^\dagger + \Delta h^\dagger , \\
\Delta h^\sharp &=& - 8 \mu \Delta \mu h^\dagger + 4(1-3\mu) \Delta h^\dagger,
\end{eqnarray}
where $\mu = m_+ m_-/m^2$ and $\Delta \mu = (m_+ -m_-)/m$.
It is noted that we have the relation $\mu = [1-(\Delta \mu)^2]/4$.
We also find the relations with respect to the average and difference
variables of the two fluids as
\begin{eqnarray}
&&m_+ n_+ U_+^\mu +  m_- n_- U_-^\mu = \rho U^\mu -\frac{1}{e} (m_+ - m_-) J^\mu ,\\
&&h_+ U_+^\mu U_+^\nu + h_- U_-^\mu U_-^\nu =
h^\dagger U^\mu U^\nu +  \frac{h^\ddagger}{(2ne)^2} J^\mu J^\nu
+ \frac{\Delta h^\dagger}{2ne} (U^\mu J^\nu + J^\mu U^\nu) , \\
&&m_- h_+ U_+^\mu U_+^\nu - m_+ h_- U_-^\mu U_-^\nu =
\frac{h^\ddagger}{4n^2e} \rho ( U^\mu J^\nu + J^\mu U^\nu )
+ \frac{\Delta h^\ddagger}{2mn^2} \rho^2 U^\mu U^\nu
+ \frac{m}{8n^2e^2} \Delta h^\sharp J^\mu J^\nu  .
\end{eqnarray}
Using the above variables and their relations, we derive the one-fluid 
equations from the two-fluid equations 
(\ref{4formnum}), (\ref{4formmom}) as
\begin{eqnarray}
&\partial_\nu &(\rho U^\nu) = 0 , \label{onefluidnum} \\
&\partial_\nu & \left [ 
h^\dagger U^\mu U^\nu + \frac{h^\ddagger}{(2ne)^2} J^\mu J^\nu 
+ \frac{\Delta h^\dagger}{2ne} (U^\mu J^\nu + J^\mu U^\nu ) \right ]
= -\partial^\mu p + J^\nu {F^\mu}_\nu ,  \\
\frac{1}{2ne}  & \partial_\nu & \left [  
\frac{h^\ddagger}{2ne} (U^\mu J^\nu + J^\mu U^\nu ) + \Delta h^\dagger U^\mu U^\nu
+ \frac{\Delta h^\sharp}{(2ne)^2} J^\mu J ^\nu \right ] \nonumber \\
&=& \frac{1}{2ne} \partial^\mu (\Delta \mu p - \Delta p) +
\left ( U^\nu - \frac{\Delta \mu}{ne} J^\nu \right) {F^\mu}_\nu + \frac{R^\mu}{ne} ,
\label{onefluidohm}
\end{eqnarray}
where $p=p_+ + p_-$, $\Delta p=p_+ - p_-$, and
\begin{equation}
R^\mu = - \eta n e \sqrt{-M} \left [ J^\mu - \frac{Q}{M} (1 + \Theta) U^\mu \right ]. 
\end{equation}
Here, $M=U^\nu U_\nu$, $Q=U^\nu J_\nu$, and $\Theta$ is the thermal
energy exchange rate from negatively charged fluid to the positive fluid
(see Appendix \ref{appenda}). In the electron-ion plasma case, we usually assume
that the energy exchange due to the friction between the two fluids is not 
significant: $\Theta =0$. 
On the other hand, in the pair plasma case, since the energy
of electron fluid and positron fluid is exchanged frequently, we set
\begin{equation}
\Theta = \frac{2m^2}{e^2} \frac{Q^2-MW}{(2mnM)^2-(mQ/e)^2} \theta  , 
\end{equation}
where $W=J_\nu J^\nu$ and $\theta$ is the redistribution 
coefficient of the thermalized energy due to the friction 
of the two fluids (see Appendix \ref{appenda}).

\section{Propagation of electromagnetic waves and causality
\label{sec3}}

With respect to the one-fluid equations of a pair plasma,
Koide (2008) indicated that causality with the electromagnetic
wave propagation breaks down 
if the electric resistivity of the pair plasma exceeds a critical value. 
To clarify the causal condition on the one-fluid equations, we investigate
electromagnetic waves in unmagnetized, uniform plasma 
using the linear analysis of the one-fluid equations
(\ref{onefluidnum})--(\ref{onefluidohm}) and the Maxwell equations
(\ref{4formfar}) and (\ref{4formamp}).
We write the variables of the uniform, rest plasma and uniform magnetic field as
$\rho = \bar{\rho}$, $p=\bar{p}$, $h^\dagger = \bar{h}$, $\VEC{U}=\VEC{0}$, 
$\VEC{B}=\VEC{0}$, and $\VEC{E}=\VEC{0}$.
When the perturbations of the variables are written by
$\tilde{\rho} =\rho - \bar{\rho}$, $\tilde{p}=p - \bar{p}$, 
$\hat{\tilde{h}} = h^\dagger - \bar{h}$, $\tilde{\VEC{U}}=\VEC{U}$, 
$\tilde{\VEC{B}}=\VEC{B}$, and $\tilde{\VEC{E}}=\VEC{E}$,
we have the linearized equations in the three-vector form,
\begin{eqnarray}
\frac{\partial}{\partial t} \tilde{\rho} 
+ \bar{\rho} \nabla \cdot \tilde{\VEC{U}} &=& 0  , \label{linearnum} \\
\frac{\partial}{\partial t} (\tilde{h} - \tilde{p})
&=& - \bar{h} \nabla \cdot \tilde{\VEC{U}} , \\
\bar{h} \frac{\partial}{\partial t} \tilde{\VEC{U}}  &=& - \nabla \tilde{p} , \\
\mu \frac{\bar{h}}{\bar{q}} \frac{\partial}{\partial t} \tilde{\VEC{K}} 
&=& \frac{1}{2\bar{q}} \nabla \cdot (\Delta \mu \tilde{p} - \Delta \tilde{p})
+\tilde{\VEC{E}} - \eta \tilde{\VEC{J}}  ,  
\label{linearohm} \\
\nabla \cdot \tilde{\VEC{E}} &=& \tilde{\rho}_{\rm e} , \\
\nabla \cdot \tilde{\VEC{B}} &=& 0  , \\
\frac{\partial}{\partial t} \tilde{\VEC{B}} &=& - \nabla \times \tilde{\VEC{E}} ,\\
\tilde{\VEC{J}} + \frac{\partial}{\partial t} \tilde{\VEC{E}} 
&=&  \nabla \times \tilde{\VEC{B}} , \label{linearamp}
\end{eqnarray}
where $q=ne$ is the ``latent" charge density, 
$\VEC{K}=\VEC{J}/(ne)$, $\bar{q}=\bar{n}e$, and
$\tilde{\VEC{K}}=\tilde{\VEC{J}}/(\bar{n}e)$.
Hereafter, we note an equilibrium variable by a bar, and perturbation
by a tilde.
We assume the resistivity is uniform and constant.
We also assume perturbation of any variables $\tilde{A}$ is proportional
to $\exp(i \VEC{k} \cdot \VEC{r} - i\omega t) = \exp(i \eta_{\mu\nu}k^\mu x^\nu)$,
where $k^\mu = (\omega, \VEC{k})$ is the constant covariant vector called
the wave number four-vector. Furthermore, for simplicity, we investigate the 
transverse modes of the electromagnetic waves in non-magnetized plasma, 
thus we set
\[
\tilde{\VEC{E}}, \verb! ! \tilde{\VEC{B}}, \verb! ! \tilde{\VEC{U}} \perp \VEC{k}. 
\]
Then we have following linearized equations,
\begin{eqnarray}
\tilde{\rho} = 0 &,& \hspace{1cm} \tilde{h} - \tilde{p} = 0, \\
-i \omega \bar{h} \tilde{\VEC{U}} &=& \VEC{0} ,  \\
-i \omega \mu \frac{\bar{h}}{\bar{q}^2} \tilde{\VEC{J}} 
&=& \tilde{\VEC{E}} - \eta \tilde{\VEC{J}} ,  \\
-i \omega \tilde{\VEC{B}} &=& -i \VEC{k} \times \tilde{\VEC{E}}  ,\\
\tilde{\VEC{J}} -i \omega \tilde{\VEC{E}} 
&=& i \VEC{k} \times \tilde{\VEC{B}}  .
\end{eqnarray}
These equations yield
\begin{eqnarray}
\left ( \eta - i\omega \mu \frac{\bar{h}}{\bar{q}^2} \right ) \tilde{\VEC{J}} 
&=&  \tilde{\VEC{E}}  ,\\
i \omega \tilde{\VEC{J}} + \omega^2 \tilde{\VEC{E}} 
&=& k^2 \tilde{\VEC{E}} .
\end{eqnarray}
Then we obtain the dispersion relation
\begin{equation}
\left ( \eta - i \mu \frac{\bar{h}}{\bar{q}^2} \omega \right ) 
(k^2-\omega^2) = i \omega.
\end{equation}
When we write $\hat{k}=k/(\eta \omega_{\rm rp}^2)$ and 
$\hat{\omega}=\omega/(\eta \omega_{\rm rp}^2)$ using 
$\omega_{\rm rp}^2 = \bar{q}^2/(\mu \bar{h})$, we have
\begin{equation}
H(\hat{k}^2 - \hat{\omega}^2) ( 1 - i\hat{\omega}) = i \hat{\omega} ,
\label{dispersion}
\end{equation}
where $H=(\eta \omega_{\rm rp})^2=(\eta \bar{q})^2/(\mu \bar{h})$.
This equation is mathematically identical to the dispersion relation of 
electromagnetic wave in the resistive pair plasma \cite{koide08b}.
The detailed analysis of the dispersion relation (\ref{dispersion})
performed by Koide (2008) shows that the group velocity of the electromagnetic
wave is smaller than the speed of light when $H < 2$, while it 
is larger than the light speed (unphysical) when $H > 3$.
That is, $H>3$ is forbidden, while the details in the range $2 < H < 3$
are shown in Koide (2008).
Consequently, the condition of the causal one-fluid equations is given by $H<2$, 
that is, 
\begin{equation}
\eta < \frac{\sqrt{2}}{\omega_{\rm rp}}  .
\end{equation}

\section{Generalized RMHD equations for electron-ion and pair plasmas
\label{sec4}}

For simplicity, we set the assumptions,
\begin{equation}
\Delta h^\dagger = 0, \hspace{1cm} M=U_\nu U^\nu = -1.
\end{equation}
When we write $h=h^\dagger$, we find
\begin{equation}
h^\ddagger = 4 \mu h, \hspace{1cm} \Delta h^\ddagger = 2 \Delta \mu h,
\hspace{1cm} \Delta h^\sharp = - 8 \mu \Delta \mu h.
\end{equation}
From equations (\ref{onefluidnum})--(\ref{onefluidohm}) and 
the above assumptions, we obtain
\begin{eqnarray}
&&\partial_\nu (\rho U^\nu) = 0 , \label{rmhdnum} \\
&&\partial_\nu  \left [ 
h \left (U^\mu U^\nu + \frac{\mu}{q^2} J^\mu J^\nu \right ) \right ]
= -\partial^\mu p + J^\nu {F^\mu}_\nu ,  \label{rmhdmom} \\
&&\frac{1}{q}   \partial_\nu  \left [  
\frac{\mu h}{q} (U^\mu J^\nu + J^\mu U^\nu ) 
- \frac{2\mu \Delta \mu h}{q^2} J^\mu J^\nu \right ] \nonumber \\
&&= \frac{1}{2q} \partial^\mu (\Delta \mu p - \Delta p) +
\left ( U^\nu - \frac{\Delta \mu}{q} J^\nu \right) {F^\mu}_\nu   
- \eta [J^\mu + Q (1+\Theta) U^\mu] .
\label{rmhdohm}
\end{eqnarray}
These equations with Maxwell equations (\ref{4formfar}) and (\ref{4formamp}) 
can be regarded to satisfy causality
when $H = (\eta q)^2/(\mu h) < 2$ everywhere.
We call equations (\ref{rmhdnum})--(\ref{rmhdohm}) with $H \la 1$
the ``generalized RMHD equations."

The generalized RMHD equations for the electron-ion plasma are given
by the limit $m \approx m_+ \gg m_-$ and an assumption $\theta =0$ as
\begin{eqnarray}
\partial_\nu (\rho U^\nu) &=& 0 , \label{normalrmhdnum} \\
\partial_\nu  ( h U^\mu U^\nu )
&=& -\partial^\mu p + J^\nu {F^\mu}_\nu , \label{normalrmhdmom} \\
\frac{m_-}{mq}   \partial_\nu  \left [  
\frac{h}{q} (U^\mu J^\nu + J^\mu U^\nu ) \right ] 
&=& \frac{1}{q} \partial^\mu p_- +
\left ( U^\nu - \frac{1}{q} J^\nu \right) {F^\mu}_\nu   
- \eta [J^\mu + Q U^\mu]  .
\label{normalrmhdohm}
\end{eqnarray}
These equations (\ref{normalrmhdnum})--(\ref{normalrmhdohm}) 
are similar to the standard RMHD equations
\begin{eqnarray}
\partial_\nu (\rho U^\nu) &=& 0 , \label{standrmhdnum} \\
\partial_\nu  ( h U^\mu U^\nu ) &=& -\partial^\mu p + J^\nu {F^\mu}_\nu ,  
\label{standrmhdmom} \\
U^\nu {F^\mu}_\nu   &=& \eta [J^\mu + Q U^\mu]  .
\label{standrmhdohm}
\end{eqnarray}
Especially, the equations with
respect to the mass density $\rho$ (equations (\ref{normalrmhdnum}) and 
(\ref{standrmhdnum})) and the momentum density (equations (\ref{normalrmhdmom}) 
and (\ref{standrmhdmom})) are identical except for the second term
of the left-hand side of equation (\ref{rmhdmom}).
The difference between the two sets of equations are found only
in the Ohm's laws (\ref{normalrmhdohm}) and (\ref{standrmhdohm}).
In the standard RMHD equations, inertia of current density, 
thermal electromotive force, and the Hall effect are ignored.

The generalized RMHD equations for a pair plasma are given
by setting $m_+ = m_- = m_{\rm e}$ ($m_{\rm e}$ is the mass of
an electron) as
\begin{eqnarray}
&&\partial_\nu (\rho U^\nu) = 0 , \label{pairrmhdnum} \\
&&\partial_\nu  \left [ 
h \left (U^\mu U^\nu + \frac{1}{(2q)^2} J^\mu J^\nu \right ) \right ]
= -\partial^\mu p + J^\nu {F^\mu}_\nu ,  \label{pairrmhdmom} \\
&&\frac{1}{4q}   \partial_\nu  \left [  
\frac{h}{q} (U^\mu J^\nu + J^\mu U^\nu ) \right ]
= - \frac{1}{2q} \partial^\mu \Delta p + U^\nu {F^\mu}_\nu   
- \eta [J^\mu + Q (1+\Theta) U^\mu].
\label{pairrmhdohm}
\end{eqnarray}
In this pair plasma case, the Hall term disappears, while the current inertia effect
is found in the equation of motion (\ref{pairrmhdmom}).
The thermal electromotive force is different from that of electron-ion plasma.
It is also noted that $\Theta$ is not negligible.

In this section, we proposed a set of generalized RMHD equations
(\ref{rmhdnum})--(\ref{rmhdohm}), the electron-ion plasma RMHD equations 
(\ref{normalrmhdnum})--(\ref{normalrmhdohm}),
and the pair plasma RMHD equations (\ref{pairrmhdnum})--(\ref{pairrmhdohm}).
The comparison between these sets of equations and the standard
RMHD equations (\ref{standrmhdnum})--(\ref{standrmhdohm}) reveals 
the unique properties of the electron-ion and pair plasmas.
We examine these unique properties in the following sections.

\section{Linear analyses of RMHD waves
\label{sec5}}

We newly proposed a set of generalized RMHD equations
(\ref{rmhdnum})--(\ref{rmhdohm}) which are applicable to electron-ion and pair
plasmas in the previous section.
In this section, we show the linear analyses of these equations
concerning various plasma waves
to show the reasonable property of the generalized RMHD equations 
and to clarify the distinct properties of the electron-ion and pair plasmas.
First, we derive the general dispersion relations with respect to 
the generalized RMHD equations, and next we reveal the unique nature
of the electron-ion and pair plasmas, respectively. 

We investigate waves propagating in a uniform, rest plasma
and a uniform magnetic field. Linearized equations of perturbations,
$\tilde{\rho} =\rho - \bar{\rho}$, $\tilde{p}=p - \bar{p}$, 
$\tilde{h} = h^\dagger - \bar{h}$, $\tilde{\VEC{U}}=\VEC{U}$, 
$\tilde{\VEC{B}}=\VEC{B} - \bar{\VEC{B}}$, and $\tilde{\VEC{E}}=\VEC{E}$,
are given by 
\begin{eqnarray}
\frac{\partial}{\partial t} \tilde{\rho} 
+ \bar{\rho} \nabla \cdot \tilde{\VEC{U}} &=& 0  , \label{linrmhdnum} \\
\bar{h} \frac{\partial}{\partial t} \tilde{\VEC{U}}  &=& - \nabla \tilde{p}
+ \tilde{\VEC{J}} \times \bar{\VEC{B}} , \\
\mu \frac{\bar{h}}{\bar{q}} \frac{\partial}{\partial t} \tilde{\VEC{K}} 
&=& \frac{1}{2\bar{q}} \nabla \cdot (\Delta \mu \tilde{p} - \Delta \tilde{p})
+(\tilde{\VEC{U}} - \Delta \mu \tilde{\VEC{K}} ) \times \bar{\VEC{B}}
+\tilde{\VEC{E}} - \eta \tilde{\VEC{J}}  ,  
\label{linrmhdohm} \\
\nabla \cdot \tilde{\VEC{E}} &=& \tilde{\rho}_{\rm e} , \\
\nabla \cdot \tilde{\VEC{B}} &=& 0  , \\
\frac{\partial}{\partial t} \tilde{\VEC{B}} &=& - \nabla \times \tilde{\VEC{E}} ,\\
\tilde{\VEC{J}} + \frac{\partial}{\partial t} \tilde{\VEC{E}} 
&=&  \nabla \times \tilde{\VEC{B}}  \label{linrmhdamp} .
\end{eqnarray}
These equations are closed with the equation of state (EoS),
$h=h(\rho,p) = \rho \hat{h}(p/\rho)$.
The adiabatic EoS for single-component relativistic fluids,
which are in thermal equilibrium, has been known, and is given by
\begin{equation}
\frac{h_\pm}{\rho_\pm} = \frac{K_3(\rho_\pm/p_\pm)}{K_2(\rho_\pm/p_\pm)}
\equiv \hat{h} (p_\pm/\rho_\pm)
\label{chandragynge}
\end{equation}
\cite{chandrasekhar38,synge57}. Here, $K_2$ and $K_3$ are the
modified Bessel functions of the second kind of order two and three,
respectively. 
Useful approximation of the EoS 
\begin{equation}
\hat{h}_{\rm RC} (\hat{p}) = 1 + \hat{p} \left ( 
1 + \frac{3 \hat{p} + 3}{3 \hat{p} + 2}
\right )
\label{hryu}
\end{equation}
was proposed by Ryu et al. (2006), where $\hat{p} = p/\rho$.
This expression has the functional dependence similar to
$\hat{h}(\hat{p})$ of equation (\ref{chandragynge}) and replaces
complex calculations of modified Bessel functions by
algebraic calculations.
When we consider the adiabatic one-component fluid
in the rest frame, equations (\ref{4formnum}) and (\ref{4formmom})
reduce to
\begin{eqnarray}
\partial_\nu (n_\pm U_\pm^\nu) &=& 0 , \\ 
\partial_\nu (h_\pm U_\pm^\mu U_\pm^\nu)  &=& -\partial^\mu p_\pm .
\end{eqnarray}
The linearized equations of the perturbations,
$\tilde{n}_\pm = n_\pm - \bar{n}_\pm$, $\tilde{p}_\pm = p_\pm - \bar{p}_\pm$,
$\tilde{h}_\pm = h_\pm - \bar{h}_\pm$, and
$\tilde{U}_\pm^\mu = U_\pm^\mu - \bar{U}_\pm^\mu$ where
$\bar{U}_\pm^\mu = (1,0,0,0)$ are
\begin{eqnarray}
\partial_0 \tilde{n}_\pm + \bar{n}_\pm \sum_i \partial_i \tilde{U}_\pm^i &= 0, \\
\partial_0 \tilde{h}_\pm + \bar{h}_\pm \sum_i \partial_i \tilde{U}_\pm^i &= 
- \partial^0 p_\pm, \\
\bar{h}_\pm \partial_0 U_\pm^i = - \partial^i p_\pm   .
\end{eqnarray}
These equations yield
\begin{equation}
\frac{\tilde{n}_\pm}{\bar{n}_\pm} 
= \frac{\tilde{h}_\pm - \tilde{p}_\pm}{\bar{h}_\pm} .
\end{equation}
From equation (\ref{chandragynge}), we find
\begin{equation}
\frac{\bar{\rho}_\pm}{\bar{p}_\pm} \frac{\tilde{p}_\pm}{\tilde{\rho}_\pm}
=\frac{\hat{h}' \left (\frac{\bar{p}_\pm}{\bar{\rho}_\pm} \right )}
{1-\hat{h}' \left (\frac{\bar{p}_\pm}{\bar{\rho}_\pm} \right)}
\equiv \Gamma  \left (\frac{\bar{p}_\pm}{\bar{\rho}_\pm} \right) .
\label{polytropicindex}
\end{equation}
In general, $\Gamma$ is not constant and is a function of 
$\bar{p}_\pm/\bar{\rho}_\pm$. 
$\Gamma$ is called the effective adiabatic index.\footnote{The polytropic
index is given by $N=(\Gamma - 1)^{-1}$.}
When $\Gamma(\bar{p}_+/\bar{\rho}_+) \approx \Gamma(\bar{p}_-/\bar{\rho}_-)$
which is satisfied only if 
$\bar{p}_\pm \ll \bar{\rho}_\pm $ or $\bar{p}_\pm \gg \bar{\rho}_\pm $ 
\cite[e.g.,][]{ryu06},\footnote{We can check this using the approximation of 
$\hat{h}_{\rm RC}$ in equation (\ref{hryu}) easily.}
we have $(1/\bar{T}) \tilde{p} = \Gamma \tilde{\rho}$,
that is, 
\begin{equation}
\frac{\bar{\rho}}{\bar{p}} \frac{\tilde{p}}{\tilde{\rho}} =
\frac{\bar{\rho}}{\bar{p}} \left ( \overline{\frac{d p}{d \rho}} 
\right )=
\Gamma \left ( \frac{\bar{p}}{\bar{\rho}} \right ).
\end{equation}
In general, 
$(\bar{\rho}/\bar{p}) (\tilde{p}/\tilde{\rho}) = \Gamma (\bar{p},\bar{\rho})$
depends on both $\bar{\rho}$ and $\bar{p}$.

\subsection{Plasma oscillation}

To show the inclusion of plasma oscillation as a fundamental mode
of the generalized RMHD equations (\ref{rmhdnum})--(\ref{rmhdohm}) with
the Maxwell equations
(\ref{4formfar}), (\ref{4formamp}), we simply assume 
\begin{equation}
\eta = 0, \bar{B}=0, \bar{p}= 0.
\end{equation}
Because $\bar{p}=0$, according to equation (\ref{polytropicindex}), we have
$\tilde{p}=\Delta \tilde{p} = 0$. The linearized equations
(\ref{linrmhdnum})--(\ref{linrmhdamp}) become
\begin{eqnarray}
\mu \frac{\bar{h}}{\bar{q}} \frac{\partial}{\partial t} \tilde{K}^i &=& E^i ,\\
\tilde{J}^i + \frac{\partial}{\partial t} \tilde{E}^i &=& 0 .
\end{eqnarray}
Immediately, we have the dispersion relation
\begin{eqnarray}
\omega^2 = \omega_{\rm rp}^2  ,
\end{eqnarray}
where $\omega_{\rm rp}^2 = \bar{q}^2/(\mu \bar{h})$.
This mode is recognized as the plasma oscillation, because $\omega_{\rm rp}$
is identical to the plasma frequency $\omega_{\rm p}=\sqrt{\bar{n}e^2/(\mu m)}$
when $\bar{p}=0$: for the electron-ion plasma, the plasma frequency is 
$\omega_{\rm rp}= \sqrt{\bar{n}e^2/m_{\rm e}}$, and in the pair plasma case,
$\omega_{\rm rp}= \sqrt{2\bar{n}e^2/m_{\rm e}}$.
As we will show in next subsection, $\omega_{\rm rp}$ is the
plasma frequency when the pressure is relativistically high.

\subsection{Compressional modes}

In this subsection, we derive a dispersion relation of a longitudinal oscillation 
modes ($\tilde{\VEC{U}} \parallel \VEC{k}$) in an unmagnetized, rest plasma
with uniform, finite pressure $\bar{p}$. For simplicity, we assume the
temperatures of the two fluids are the same: $\bar{T}=\bar{T}_+=\bar{T}_-$.
Using equation (\ref{averho}) and 0-th component of
equation (\ref{ave4cur}), we have
$\gamma n_\pm = (\gamma \rho \pm m_\mp \rho_{\rm e}/e)/m$ when
$\gamma = \gamma_+ \approx \gamma_-$. Using these equations, we have
\begin{equation}
\frac{\partial \gamma n}{\partial t} = - \frac{2}{m} \nabla \cdot (\rho \VEC{U})
+ \frac{\mu}{e} (\nabla \cdot \VEC{J}).
\end{equation}
In the present non-relativistic case, setting $\gamma=1$, we have
\begin{equation}
\frac{\partial n}{\partial t} = - \frac{2}{m} \nabla \cdot (\rho \VEC{U})
+ \frac{\mu}{e} (\nabla \cdot \VEC{J}).
\end{equation}
If $\Gamma(\bar{p}_+/\bar{\rho}_+) = \Gamma(\bar{p}_-/\bar{\rho}_-) 
\equiv \Gamma$ is uniform and constant, we obtain
\begin{eqnarray}
\frac{\partial \tilde{p}}{\partial t} &=& 
- \frac{2 \Gamma \bar{T}}{m} \left [ \nabla \cdot (\bar{\rho} \tilde{\VEC{U}}) 
- \frac{\Delta \mu}{2e} (\nabla \cdot \tilde{\VEC{J}}) \right ], \\
\frac{\partial}{\partial t} \Delta \tilde{p} &=& 
- \frac{\Gamma \bar{T}}{m} \nabla \cdot \VEC{J} .
\end{eqnarray}
We have the linearized equations,
\begin{eqnarray}
\bar{h} \frac{\partial}{\partial t} \tilde{\VEC{U}} &=& - \nabla \tilde{p}, \\
\frac{\mu \bar{h}}{\bar{q}^2} \frac{\partial}{\partial t} \tilde{\VEC{J}} &=& 
\frac{1}{2\bar{q}} \nabla (\Delta \mu \tilde{p} - \Delta \tilde{p} ) 
+ \tilde{\VEC{E}}, \\
\tilde{\VEC{J}} + \frac{\partial}{\partial t} \tilde{\VEC{E}} &=& \VEC{0}.
\end{eqnarray}
These equations yield
\begin{eqnarray}
\frac{\mu \bar{h}}{\bar{q}^2} \frac{\partial^2}{\partial t^2} \tilde{\VEC{J}} &=& 
-\frac{\Delta \mu \Gamma \bar{T}}{e} \nabla (\nabla \cdot \tilde{\VEC{U}})
+\frac{\Gamma \bar{T}}{2\bar{q} e} (1 + (\Delta \mu)^2) 
\nabla (\nabla \cdot \tilde{\VEC{J}}) - \tilde{\VEC{J}}, \\
\bar{h} \frac{\partial^2}{\partial t^2} \tilde{\VEC{U}} &=& 
\frac{2 \Gamma \bar{T} \bar{\rho}}{m} \nabla (\nabla \cdot \tilde{\VEC{U}}) 
-\frac{\Delta \mu \Gamma \bar{T}}{e} \nabla (\nabla \cdot \tilde{\VEC{J}}) .
\end{eqnarray}
Because $\tilde{\VEC{J}} \parallel \VEC{k}$ and 
$\tilde{\VEC{U}} \parallel \VEC{k}$ in the longitudinal modes, we obtain
\begin{eqnarray}
-\omega^2 \frac{\mu \bar{h}}{\bar{q}^2} \tilde{J}_\parallel &=&
\frac{\Delta \mu \Gamma \bar{T}}{e} k^2 \tilde{U}_\parallel
-\frac{\Gamma \bar{T}}{2\bar{q}e} (1+(\Delta \mu)^2) k^2 \tilde{J}_\parallel
- \tilde{J}_\parallel, \\
-\omega^2 \bar{h} \tilde{U}_\parallel &=&
- \frac{2 \Gamma \bar{T} \bar{\rho}}{m} k^2 \tilde{U}_\parallel
+\frac{\Delta \mu \Gamma \bar{T}}{e} k^2 \tilde{J}_\parallel ,
\end{eqnarray}
where $\tilde{J}_\parallel \equiv (\VEC{J} \cdot \VEC{k})/k$ 
and $\tilde{U}_\parallel \equiv (\VEC{U} \cdot \VEC{k})/k$, $k \ne 0$.
Then we get the following dispersion relation,
\begin{equation}
\left [ 
\omega^2 - \frac{c_{\rm s}^2}{2\mu} (1+(\Delta \mu)^2) k^2 - \omega_{\rm rp}^2
\right ]
(\omega^2 - 2 c_{\rm s}^2 k^2) =
\frac{(\Delta \mu)^2}{\mu} (c_{\rm s} k)^4   ,
\end{equation}
where $c_{\rm s}^2 = \Gamma \bar{p}/\bar{h} = (\bar{\rho}/\bar{h})(dp/d\rho)$.

In a case of an electron-ion plasma ($\Delta \mu \approx 1$, $\mu = m_{\rm e}/m \ll 1$),
when $\omega \gg c_{\rm s} k/\sqrt{\mu}$, 
the dispersion relation becomes
\begin{equation}
\omega^2 = \frac{c_{\rm s}^2}{2 \mu} k^2 + \omega_{\rm rp}^2.
\end{equation}
This expression shows the dispersion relation of the plasma oscillation
for the plasma with finite pressure.
In the case that $\omega^2 \ll \omega_{\rm rp}^2/\mu$, we have the dispersion relation
of sound waves
\begin{equation}
\omega^2 = 2 c_{\rm s}^2 k^2 .
\end{equation}
In a case of a pair plasma ($\Delta \mu = 0$, $\mu = 1/4$), we have
two modes
\begin{equation}
\omega^2 =  \omega_{\rm rp}^2 +  2 c_{\rm s}^2 k^2 ,
\end{equation}
and
\begin{equation}
\omega^2 =  2 c_{\rm s}^2 k^2  .
\end{equation}
The former is the dispersion relation of plasma oscillation and
the latter is that of the sound wave.

We investigate the compressional wave of the magnetized plasma.
When $\VEC{k} \parallel \bar{\VEC{B}}$ and 
$\tilde{\VEC{U}} \parallel \bar{\VEC{B}}$, 
the dispersion relation is the
same as that of the non-magnetized plasma wave. Then, we investigate 
the case that
$\VEC{k} \perp \bar{\VEC{B}}$, $\VEC{k} \perp \tilde{\VEC{B}}$, 
$\VEC{k} \parallel \tilde{\VEC{U}}$,
and $\rho_{\rm e} = 0$ in this paragraph. Because $\rho_{\rm e}=0$, we have
$\Delta p = 0$. In the case that 
$k \ll \omega_{\rm rp}$ and $\omega \ll \omega_{\rm rc}$, 
the left-hand side, first term and the Hall term
of the right-hand side of the Ohm's law (\ref{linrmhdohm}) are negligible.
The validity of this assumption is checked bellow. 
Here, we defined a characteristic frequency of the Larmor rotation of an
averaged particle of the two-fluid model by
\begin{equation}
\omega_{\rm rc} = \frac{e \bar{B}}{m} \frac{\bar{\rho}}{\bar{h}}.
\end{equation}
If we set the pressure to be zero, $\omega_{\rm rc}$ reduces to the cyclotron 
frequency of the charged particle with mass $m$ and charge $e$
in the magnetic field $\bar{B}$, $\omega_{\rm c}=e\bar{B}/m$.
Using linearized equations (\ref{linrmhdnum})--(\ref{linrmhdamp}), 
we have the dispersion relation,
\begin{equation}
\omega^2 = v_{\rm f}^2 k^2 , \label{disprelfast}
\end{equation}
where $v_{\rm f}^2 = (\Gamma \bar{p} + \bar{B}^2)/(\bar{h} + \bar{B}^2)$.
This mode shows the dispersion relation of the fast wave.
It is also noted that $v_{\rm f} < 1$.
Below, we check the validity of the assumption 
$k \ll \omega_{\rm rp}$ and $\omega \ll \omega_{\rm rc}$ that we used.
Using the Ohm's law (\ref{linrmhdohm}) with this assumption and 
the dispersion relation (\ref{disprelfast}),
the left-hand side of the Ohm's law (\ref{linrmhdohm}) becomes
\begin{equation}
i \omega \frac{\mu \bar{h}}{\bar{q}^2} \tilde{\VEC{J}} =
\left ( \frac{k}{\omega_{\rm rp}} \right )^2 (v_{\rm f}^2 -1)
\tilde{\VEC{U}} \times \bar{\VEC{B}} .
\end{equation}
This clearly shows that the term of the left-hand side 
can be ignored in equation (\ref{linrmhdohm})
when $k \ll \omega_{\rm rp}$.
The absolute value of the summation of the thermal electromotive force term
and the Hall term of the Ohm's law (\ref{linrmhdohm}) is estimated by
\begin{eqnarray}
&&\left | \frac{1}{2\bar{q}} i \VEC{k} \Delta \mu \tilde{p} - \Delta \mu 
\tilde{\VEC{K}} \times \bar{\VEC{B}} \right | 
= \left | i \frac{\Delta \mu}{\bar{q}} \omega \tilde{\VEC{U}}
\left [ \left (\frac{\Gamma \bar{p}}{2} + \bar{B}^2 \right ) \frac{1}{v_{\rm f}^2} 
- \bar{B}^2 \right ] \right |   \\
&& = \frac{\Delta \mu}{\bar{q}} \omega \tilde{U} \left |
\left (\frac{\Gamma \bar{p}}{2} + \bar{B}^2 \right ) \frac{1}{v_{\rm f}^2} 
- \bar{B}^2 \right |    
\leq \frac{\Delta \mu}{\bar{q}} \omega \bar{h} \tilde{U}   .
\end{eqnarray}
This is negligible compared to 
$| \tilde{\VEC{U}} \times \bar{\VEC{B}} | \sim \tilde{U} \bar{B}$ 
because the ratio of the two terms is $\sim \Delta \mu \omega / \omega_{\rm rc}$ 
when $\omega \ll \omega_{\rm rc}$.

\subsection{Transverse wave propagating along magnetic field}

We investigate transverse waves propagating through the ideal
MHD plasma along the magnetic field lines,
\begin{equation}
\bar{\VEC{B}} \parallel \VEC{k}, \hspace{1cm}
\tilde{\VEC{E}}, \verb! ! \tilde{\VEC{B}}, \verb! ! \tilde{\VEC{U}} 
\parallel \VEC{k}, \hspace{1cm} \eta = 0.
\end{equation}
We assume any perturbation $\tilde{A}$ is proportional to 
$\exp(i \VEC{k} \cdot \VEC{r}) \tilde{A} (t)$. 
The linearized equations become
\begin{eqnarray}
\bar{h} \frac{\partial}{\partial t} \tilde{\VEC{U}} &=& 
\tilde{\VEC{J}} \times \bar{\VEC{B}} ,\\
\frac{\mu \bar{h}}{\bar{q}} \frac{\partial}{\partial t} \tilde{\VEC{J}}
&=& (\bar{q} \tilde{\VEC{U}} - \Delta \mu \tilde{\VEC{J}}) \times \bar{\VEC{B}}
+ \bar{q} \tilde{\VEC{E}}  , \\
i \VEC{k} \cdot \tilde{\VEC{E}} &=& 0 ,\\
i \VEC{k} \cdot \tilde{\VEC{B}} &=& 0 ,\\
\frac{\partial}{\partial t} \tilde{\VEC{B}} &=& 
- i \VEC{k} \times \bar{\VEC{E}} ,\\
\tilde{\VEC{J}} + \frac{\partial}{\partial t} \tilde{\VEC{E}} &=& 
 i \VEC{k} \times \bar{\VEC{B}} .
\end{eqnarray}
From these linearized equations, we have
\begin{equation}
\frac{\mu \bar{h}}{\bar{q}} \frac{\partial^2}{\partial t^2} \Box \tilde{\VEC{J}}
= - \frac{\bar{q}}{\bar{h}} (\Box \tilde{\VEC{J}}) \bar{B}^2 
- \Delta \mu \frac{\partial}{\partial t} (\Box \tilde{\VEC{J}}) \times \bar{\VEC{B}}
+ \bar{q} \frac{\partial^2}{\partial t^2} \tilde{\VEC{J}}  , 
\end{equation}
where $\Box = \partial_\nu \partial^\nu$. When we set
\begin{equation}
\frac{\partial \tilde{\VEC{J}}}{\partial t} = i \omega \tilde{\VEC{J}}
+ \VEC{\Omega} \times \tilde{\VEC{J}}  ,
\end{equation}
we obtain the following dispersion relations,
\begin{eqnarray}
\frac{\bar{q} \bar{B}^2}{\bar{h}} k'^2 = \bar{q} \omega'^2
&+& \frac{\mu \bar{h}}{\bar{q}} (\omega'^2 k'^2 - 4\omega^2 \Omega^2)
-\Delta \mu (k'^2 - 2 \omega^2) \Omega_\parallel \bar{B}  , \label{dispreltra1} \\
2 \Omega_\parallel \left [ \frac{\mu \bar{h}}{\bar{q}} (k'^2 - \omega'^2) 
+ \bar{q} + \frac{\bar{q} \bar{B}^2}{\bar{h}}
\right ] &=& \Delta \mu (k'^2 -2 \Omega^2) \bar{B}, \label{dispreltra2} 
\end{eqnarray}
where $k'^2=k^2-\omega^2-\Omega^2$ and $\omega'^2=\omega^2+\Omega^2$.

Let us consider the case of pair plasma ($\mu = 1/4$, $\Delta \mu =0$).
In this case, we have $\Omega = 0$, $\omega'^2=\omega$, $k'^2=k^2-\omega^2$, 
and the dispersion relation becomes
\begin{equation}
\left [ \frac{\bar{B}^2}{\bar{h}}  
+ \left ( \frac{\omega}{\omega_{\rm rp}} \right )^2 \right ] 
(k^2 - \omega^2) = \omega^2  .
\end{equation}
When $\omega_{\rm rp}^2 \gg \omega^2$, we have 
\begin{equation}
\omega^2 = \frac{u_{\rm A}^2}{1+u_{\rm A}^2} k^2  ,
\end{equation}
where $u_{\rm A} = \sqrt{\bar{B}^2/\bar{h}}$ is the Alfven four-velocity.
This shows the dispersion relation of the relativistic Alfven wave.

For the electron-ion plasma case ($\mu \approx m_{\rm e}/m \ll 1$, 
$\Delta \mu \approx 1$),
we arrange the dispersion equations (\ref{dispreltra1}), (\ref{dispreltra2}) as
\begin{eqnarray}
u_{\rm A}^2 k'^2 &=& \omega'^2 
- (k'^2 - 2 \omega^2) \frac{\omega_{\rm rc}}{\omega_{\rm rp}^2} \Omega_\parallel , 
\label{disprelnoralf} \\
2 (1+u_{\rm A}^2) \Omega_\parallel &=& (k^2-2\Omega^2) 
\frac{\omega_{\rm rc}}{\omega_{\rm rp}^2}  .
\end{eqnarray}
It is noted that there is the relation between $\omega_{\rm rc}$ and
$\omega_{\rm rp}$,
\begin{equation}
\frac{\omega_{\rm rc}}{\omega_{\rm rp}^2} = \frac{\bar{B}}{\bar{q}} .
\end{equation}
When $\Omega \la k$, $k \ll \omega_{\rm rp}$, $\omega_{\rm rc} \ll \omega_{\rm rp}$, 
then $\Omega_\parallel \ll k$ and $k'^2 = k^2$, thus we have
\begin{equation}
\Omega_\parallel = \frac{\omega_{\rm rc}}{2 \omega_{\rm rp}^2} 
\frac{k^2}{1+ u_{\rm A}^2}  . \label{alfw2bef}
\end{equation}
Substituting this equation to equation (\ref{disprelnoralf}), we obtain
\begin{equation}
u_{\rm A}^2 (k^2 - \omega^2) = \omega^2
\left [ 1 + \frac{3}{2} 
\left ( \frac{\omega_{\rm rc} k}{\omega_{\rm rp}^2} \right )^2
\frac{1}{1+u_{\rm A}^2} \right ] - \frac{1}{4}
\left ( \frac{\omega_{\rm rc}}{\omega_{\rm rp}^2} k^2 \right )^2  .
\end{equation}
When $\omega_{\rm rc} \ll \omega_{\rm rp}$, $k \ll \omega_{\rm rp}$, 
we finally have
\begin{equation}
\omega^2 = u_{\rm A}^2 (k^2 - \omega^2) . \label{alfw1bef}
\end{equation}
Equations (\ref{alfw2bef}), (\ref{alfw1bef})
yield the dispersion relation of the Alfven wave in the electron-ion plasma case,
\begin{eqnarray}
\omega^2 &=& \frac{u_{\rm A}^2}{1+ u_{\rm A}^2} k^2 ,\\
\Omega_\parallel &=& \frac{\omega_{\rm rc}}{2(1+u_{\rm A}^2) \omega_{\rm rp}^2} k^2 .
\end{eqnarray}
It is noted that the polarization of the Alfven wave rotates, while the
rotation angular velocity $\Omega_\parallel$ is very small.

We examine the electromagnetic waves propagating along the
magnetic field lines.
From equations  (\ref{dispreltra1}), (\ref{dispreltra2}), we have
\begin{eqnarray}
\omega'^2 k'^2 - 4\omega^2 \Omega^2 + \omega_{\rm rp}^2 \omega'^2
-\mu \omega_{\rm rc}^2 k'^2 &=& \Delta \mu \omega_{\rm rc}
(k'^2 - 2\omega^2) \Omega_\parallel , \\
2 \Omega_\parallel [ k'^2 - \omega'^2 + \omega_{\rm rp}^2
+ \mu \omega_{\rm rc}^2 ] &=& \Delta \mu \omega_{\rm rc} 
(k'^2 - 2\Omega^2)  .
\end{eqnarray}
When $\omega \gg \omega_{\rm rp}, \omega_{\rm rc}$, the medium
of the electromagnetic wave becomes nearly vacuum, and we assume
\begin{equation}
\omega \gg k' \gg \Omega .
\end{equation}
Then using $\omega' = \omega$, we have
\begin{eqnarray}
\omega^2 (k'^2 + \omega_{\rm rp}^2 )
&=& (\Delta \mu)^2 \omega_{\rm rc}^2 k'^2  , \\
-2 \Omega_\parallel \omega^2 &=& \Delta \mu \omega_{\rm rc} k'^2  .
\end{eqnarray}
Thus, we have the dispersion relations,
\begin{eqnarray}
\omega^2 &=& k^2 + \omega_{\rm rp}^2  ,\\
\Omega_\parallel &=& \Delta \mu \frac{\omega_{\rm rc} \omega_{\rm rp}^2}{2 \omega^2} .
\end{eqnarray}
The finite $\Omega_\parallel$ shows the Faraday rotation of the electromagnetic
wave along the magnetic field line in the electron-ion plasma.
In the pair plasma case ($\Delta \mu = 0$), no rotation of polarization is found.

\section{Unique nature of electron-ion and pair plasmas
\label{sec6}}

The differences between the standard RMHD equations and generalized
RMHD equations of electron-ion plasmas and pair plasmas are only following five points.
To find out these points, we should just consider the terms with $\mu$ and $\Delta \mu$
in equations (\ref{rmhdnum})--(\ref{rmhdohm}).
\begin{enumerate}
\item Second term of left-hand side of equation of motion (\ref{rmhdmom}):
This term is due to the momentum of current.
In the pair plasma case, it is significant, while it is negligible
in an electron-ion plasma.
\label{pecu1}

\item The time-derivative term of the Ohm's law (\ref{rmhdohm}):
This is due to the inertia effect of the current. 
We can ignore it only when the electric resistivity is zero 
in an electron-ion plasma. When the resistivity is finite, we cannot ignore it 
because of causality.
\label{pecu2}

\item The second term with respect to $J^\mu$ of the right-hand side
of the Ohm's law (\ref{rmhdohm}):
This is the Hall term. In the pair plasma case, it disappears.
\label{pecu3}

\item The term with respect to pressure gradient of the right-hand side
of the Ohm's law (\ref{rmhdohm}):
This is the term of the thermal electromotive force.
In a pair plasma, it influences through $\Delta p$, while in
an electron-ion plasma through $p_-$.
\label{pecu4}

\item The term proportional to $\Theta$ in the Ohm's law (\ref{rmhdohm}):
This is due to the difference of the thermalized energy
exchange between the electron--positron pair and 
electron--ion pair during the friction.
\label{pecu5}
\end{enumerate}
The item 3 was indicated by many authors \cite[e.g.,][]{blackman93,meier04} 
and the other items are explicitly described in this paper for the first time.
As shown above, it is clear that most of differences come from the 
Ohm's law (\ref{rmhdohm}).
The exception is the effect of current moment in the equation of
motion shown in the distinct property \ref{pecu1}.
The influence of the distinct property \ref{pecu1} is 
estimated by comparison between $\sqrt{\mu} J^\mu/(ne)$ and $U^\mu$,
which are on the order of the square root of two terms
in the left-hand side of equation (\ref{rmhdmom}).
This implicitly contains two ratios,
\begin{equation}
\sqrt{\mu} \frac{J^0}{ne} : 1 , \hspace{1cm}
\sqrt{\mu} \frac{|\VEC{J}|}{ne} : |\VEC{U}| 
= \sqrt{\mu} m|\VEC{U}_+ - \VEC{U}_-| : |m_+ n_+ \VEC{U}_+ + m_- n_- \VEC{U}_-  | .
\label{ratio_pecu1}
\end{equation}
To evaluate the latter ratio, we introduce the transformation
between the two frames $(t, \VEC{r})$ and $(t', \VEC{r'})$ 
whose relative velocity is $\VEC{v}_0$, like the Galilean transformation,
\begin{equation}
t' = t, \hspace{1cm}
\VEC{r}^\flat = \VEC{r} - \VEC{v}_0 t  .
\end{equation}
For any three-vector $\VEC{A}$, the vector length $|\VEC{A}^\flat|$
is equal to or smaller than the length of the vector transformed by the Lorentz
transformation, $|\VEC{A}'|$.
This is because
\[
|\VEC{A}'| = |\VEC{A}'_\parallel + \VEC{A}'_\perp|
= |\gamma_0 (\VEC{A}_\parallel - \VEC{v}_0 t) + \VEC{A}_\perp|
=\sqrt{|\gamma_0 (\VEC{A}_\parallel - \VEC{v}_0 t)|^2 + |\VEC{A}_\perp|^2}
\]
\[
\geq \sqrt{|\VEC{A}_\parallel - \VEC{v}_0 t|^2 + |\VEC{A}_\perp|^2}
=|\VEC{A}_\parallel - \VEC{v}_0 t + \VEC{A}_\perp|
=|\VEC{A}^\flat_\parallel + \VEC{A}^\flat_\perp|
=|\VEC{A}^\flat| ,
\]
where the vector with the subscript $\parallel$ ($\perp$) is its parallel 
(perpendicular) component to $\VEC{v}_0$ and $\gamma_0=(1-v_0^2)^{-1/2}$.
When we take the velocity of the center-of-mass of the two fluid
locally as $\VEC{v}_0$, that is, $\VEC{v}_0 = \VEC{U}$,
we have
\[
\Delta \VEC{U}^\flat \equiv \VEC{U}_+^\flat - \VEC{U}_-^\flat
=\frac{1}{\gamma} (\gamma_- \VEC{U}_+ - \gamma_+ \VEC{U}_-)
=\frac{\gamma_+ \gamma_-}{\gamma} (\VEC{V}_+ - \VEC{V}_-)  ,
\]
and
\[
|\Delta \VEC{U}^\flat | \leq | \VEC{U}_+' - \VEC{U}_-' |  .
\]
We can write
\begin{eqnarray}
\VEC{U}_\pm &=& \frac{1}{\rho \gamma} (\gamma_\pm \rho \VEC{U} \pm 
\gamma_\mp \rho \Delta \VEC{U}^\flat)  , \\
\VEC{K} &=& 
\frac{\rho_{\rm e}}{en} \VEC{V} + \frac{n_+ n_-}{n^2} \Delta \VEC{U}^\flat .
\end{eqnarray}
In general, since $\Delta \mu \leq 1$, the condition of the neglect of 
the current momentum is
\begin{equation}
\rho_{\rm e} \ll \frac{en}{\sqrt{\mu}}, \hspace{1cm} 
|\VEC{U}_+ - \VEC{U}_-| \ll \frac{|\VEC{U}|}{\sqrt{\mu}}.
\label{cond_pecu1}
\end{equation}
This is because
\[
\left | \frac{\VEC{J}}{ne} \right | = |\VEC{K}|
=\left | \frac{\rho_{\rm e}}{en} \VEC{V} + \frac{n_+ n_-}{n^2} \Delta \VEC{U}^\flat
\right |
\leq \frac{\rho_{\rm e}}{ne} |\VEC{V}| + \frac{n_+ n_-}{n^2} |\Delta \VEC{U}^\flat|
< \frac{\rho_{\rm e}}{ne} |\VEC{U}| + |\VEC{U}_+ - \VEC{U}_-| \ll |\VEC{U}|.
\]

The influence of the unique nature \ref{pecu3} is estimated by comparison
between $\Delta \mu J^\mu/(ne)$ and $U^\mu$ in the second term,
the right-hand side of equation (\ref{rmhdohm}).
As the above consideration, if
\begin{equation}
\rho_{\rm e} \ll \frac{en}{\Delta \mu}, \hspace{1cm} 
|\VEC{U}_+ - \VEC{U}_-| \ll \frac{|\VEC{U}|}{\Delta \mu}  ,
\label{cond_pecu3}
\end{equation}
the unique nature \ref{pecu3} is negligible. 
In the pair plasma case ($\Delta \mu =0$), the Hall effect disappears
without any restriction.

The distinct property \ref{pecu2}, the time-derivative term of the Ohm's law,
is evaluated by the ratio
\begin{equation}
\frac{\mu}{c \tau} \frac{h}{(ne)^2} U J : U^\nu {F^i}_\nu
\sim \frac{\mu}{c \tau} \frac{h}{(ne)^2} J : B  .
\label{ratio_pecu2}
\end{equation}
The ratio should be
evaluated in a more realistic case because it depends on the characteristic 
time scale $\tau$. An example on black hole magnetospheres of AGNs 
is discussed in the next section.
With respect to the unique property \ref{pecu4} of the thermal electromotive
force, the influence is estimated by the ratio
\begin{equation}
\frac{\Delta \mu p - \Delta p}{2nel} : U^\nu F_{\nu i}  ,
\end{equation}
where $l$ is the characteristic scale length.
As for the unique nature \ref{pecu5}, the term is negligible
in the electron-ion plasma case in a short time scale, because of the energy
exchange between the electron fluid and ion fluid. 
However, we cannot ignore it in the pair plasma case.
In this paper, we do not discuss the details of the last two 
distinct properties.

\section{Validity of generalized and standard RMHD equations
\label{sec7}}

In this section, we show the applicability of the generalized RMHD equations 
and evaluate the significance of the pair and electron-ion plasma 
distinct properties in a global, astrophysical situation.
The standard RMHD equations are applicable only 
when the conditions of the generalized RMHD are satisfied and the unique properties
of pair and electron-ion plasmas can be ignored (see section \ref{sec6}).


\subsection{Break-down condition of generalized RMHD equations
\label{sec7bdc}}

We consider the break-down condition of the generalized RMHD equations
(\ref{rmhdnum})--(\ref{rmhdohm}) with the Maxwell equations 
(\ref{4formfar}), (\ref{4formamp})
on the basis of the two-fluid approximation and the average procedure
(\ref{averho})--(\ref{ave4cur}). To derive the generalized RMHD equations from
the one-fluid equations, we used the assumptions $M=U_\nu U^\nu = -1$ and
$\Delta h^\dagger \ll h$. It is noted that the one-fluid equations
(\ref{onefluidnum})--(\ref{onefluidohm}) and the two-fluid equations 
(\ref{4formnum})--(\ref{4formmom}) 
are consistent within the averaging procedure (\ref{averho})--(\ref{ave4cur}). 
The condition $M=-1$ is necessary so that $U^\mu$ has meaning of the 
four-velocity, and that means the averaged Lorentz factor of the fluid
of positively charged particles and the other fluid of negatively charged
particles is non-relativistic in the center-of-mass frame because 
$M=U_\nu U^\nu = {U_\nu}' {U^\nu}' = {U_0}' {U^0}' = -(\gamma')^2 =-1$.
Here, the prime indicates a variable observed by the local center-of-mass frame
of the two fluids.
This condition is consistent with the assumption of the current-density
independency of the resistivity (see Appendix A). When the relative
velocity of the two fluids is relativistic, the frictional force of the
two fluid is not proportional to the relative velocity, and then
the resistivity depends on the current density.
This condition restricts the net charge density and current density as follows.
Using equations (\ref{ave4vel}) and (\ref{ave4cur}), we get
\begin{equation}
U_\nu U^\nu = -1 - \frac{m_+ m_-}{\rho^2}
\left [ (n_+ - n_-)^2 + \frac{1}{e^2} J_\nu J^\nu  \right ] ,
\end{equation}
by some algebraic calculations.
When $M = U_\nu U^\nu \approx 1$, we find
\[
\frac{m_+ m_-}{\rho^2}
\left [ (n_+ - n_-)^2 + \frac{1}{e^2} J_\nu J^\nu , \right ] \ll 1 .
\]
Using this condition and $J_\nu J^\nu = - \rho_{\rm e}^2 + |\VEC{J}|^2$,
we find
\[
(n_+ - n_-)^2 + \frac{1}{e^2} |\VEC{J}|^2
\ll \frac{\rho_{\rm e}^2}{e^2} + \frac{\rho^2}{m_+ m_-}
= \frac{\rho_{\rm e}^2}{e^2} + \frac{\rho^2}{m^2 \mu} ,
\]
then we have
\begin{equation}
|\VEC{J}|^2 \ll \rho_{\rm e}^2 + \frac{(en)^2}{\mu} .
\label{prim_gamma}
\end{equation}
This condition shows the current must be much smaller than
$en/\sqrt{\mu}$ or $\rho_{\rm e}$.

For an electron-ion plasma, if $\gamma_-' \ll m_+/m_-$ and $n_+ \approx n_-$, 
we have the condition $\gamma' \approx 1$.
This is because 
\[
\gamma' \approx \gamma_+' + \frac{m_-}{m_+} \gamma_-' \approx 1,
\]
since
\[
|\VEC{U}_+' | = \frac{m_- n_-}{m_+ n_+} |\VEC{U}_-' | 
\leq \frac{m_-}{m_+}  \gamma_-'
\ll 1. 
\]
For a pair plasma, we need $\gamma_\pm' \approx 1$
to keep the condition of $\gamma' \approx 1$ 
because $\gamma_+' \approx \gamma_-'$ and 
$\gamma' \approx (\gamma_+' + \gamma_-')/2 \approx \gamma_\pm'$
if $n_+ \approx n_-$.

With respect to the condition $\Delta h^\dagger \ll h^\dagger$,
in the case of the non-relativistic pressure limit $p_\pm \ll \rho_\pm$,
we find
\[
2\mu m \left ( \frac{1}{n_+} - \frac{1}{n_-} \right )
\ll \frac{m_+}{n_+} + \frac{m_-}{n_-} .
\]
This condition requires $n_+ \approx n_-$.
It is noted that this does not mean the charge neutrality 
because net charge density is given by
$\rho_{\rm e} = e(n_+ \gamma_+ - n_- \gamma_- ) \ne e(n_+ - n_-)$.
We call the condition $n_+ \approx n_-$ the ``proper charge neutrality".
In the relativistic pressure case, we find this condition yields
\[
\frac{2}{m} \left ( \frac{h_+ m_-}{n_+^2} - \frac{h_- m_+}{n_-^2} \right )
\ll \frac{h_+}{n_+^2} + \frac{h_-}{n_-^2} ,
\]
and then using specific enthalpy (\ref{chandragynge}), which is approximated by
equation (\ref{hryu}), we have
\begin{equation}
\frac{1}{n_+} \hat{h} \left ( \frac{p_+}{\rho_+} \right ) 
\approx \frac{1}{n_-} \hat{h} \left ( \frac{p_-}{\rho_-} \right ) .
\label{condrelpres}
\end{equation}
This condition is satisfied when $n_+ \approx n_-$
and  $p_+/\rho_+ \approx p_-/\rho_-$ except for a special case.

In the case of an electron-ion plasma with proper charge neutrality,
the condition $p_+/\rho_+ \approx p_-/\rho_-$ is implausible 
when the ion temperature $T_+$ is not much larger than the electron
temperature $T_-$ because it requires
$T_+ = p_+/n_+ \approx (m_+/m_-) m_- p_-/\rho_- = (m_+/m_-) T_-$
and $m_+ \gg m_-$. Then, the application condition of the generalized RMHD
equations is $p_\pm \ll \rho_\pm$. In a case of a pair plasma
with proper charge neutrality, the application condition
is $p_+/\rho_+ \approx p_-/\rho_-$ or $p_\pm \ll \rho_\pm$.
In general (both cases of the electron-ion and pair plasmas), the generalized
RMHD equations are applicable to the plasmas with the proper charge
neutrality $n_+ \approx n_-$, the non-relativistic thermal energy 
$p_\pm \ll \rho_\pm$, and the non-relativistic (non-superrelativistic
for electron-ion plasmas) relative velocity
of the two fluids $U_+' \leq U_-' \ll m_+/m_-$ if $m_+ \geq m_-$.

The premise conditions of the generalized RMHD equations are 
summarized as follows. The generalized RMHD approximation breaks
down if all of the following conditions are not satisfied.

\begin{enumerate}
\item The local relative velocity of the two fluids is non-relativistic,
$\gamma' \approx 1$. This condition comes from the condition $U_\nu U^\nu = -1$.
This condition restricts the current density as shown by equation 
(\ref{prim_gamma}). In an electron-ion plasma, it is satisfied when 
$\gamma_-' \ll m_+/m_-$ and $n_+ \approx n_-$, 
while in a pair plasma it requires $\gamma_\pm' \approx 1$.

\item The condition of the proper charge density $n_+ \approx n_+$
is required to hold the condition $\Delta h^\dagger \ll h^\dagger$.
It is noted that this condition does not mean the charge neutrality 
$\rho_{\rm e} \approx 0$. In the non-relativistic pressure case 
($p_\pm \ll \rho_\pm$), the condition $\Delta h^\dagger \ll h^\dagger$ 
is satisfied only with this condition.

\item In the relativistic pressure case, we require the condition shown by
equation (\ref{condrelpres}). In the electron-ion plasma case, this condition
yields $p_\pm \ll \rho_\pm$ except for the special case such as
$T_+ \approx (m_+/m_-) T_-$. 
In the pair plasma case, it requires $T_+ \approx T_-$ or $p_\pm \ll \rho_\pm$.
\end{enumerate}

\subsection{Unique nature significance of electron-ion/pair plasmas 
in AGN jet forming region \label{sec7bhj}}

\begin{figure}
\epsscale{.90}
\plotone{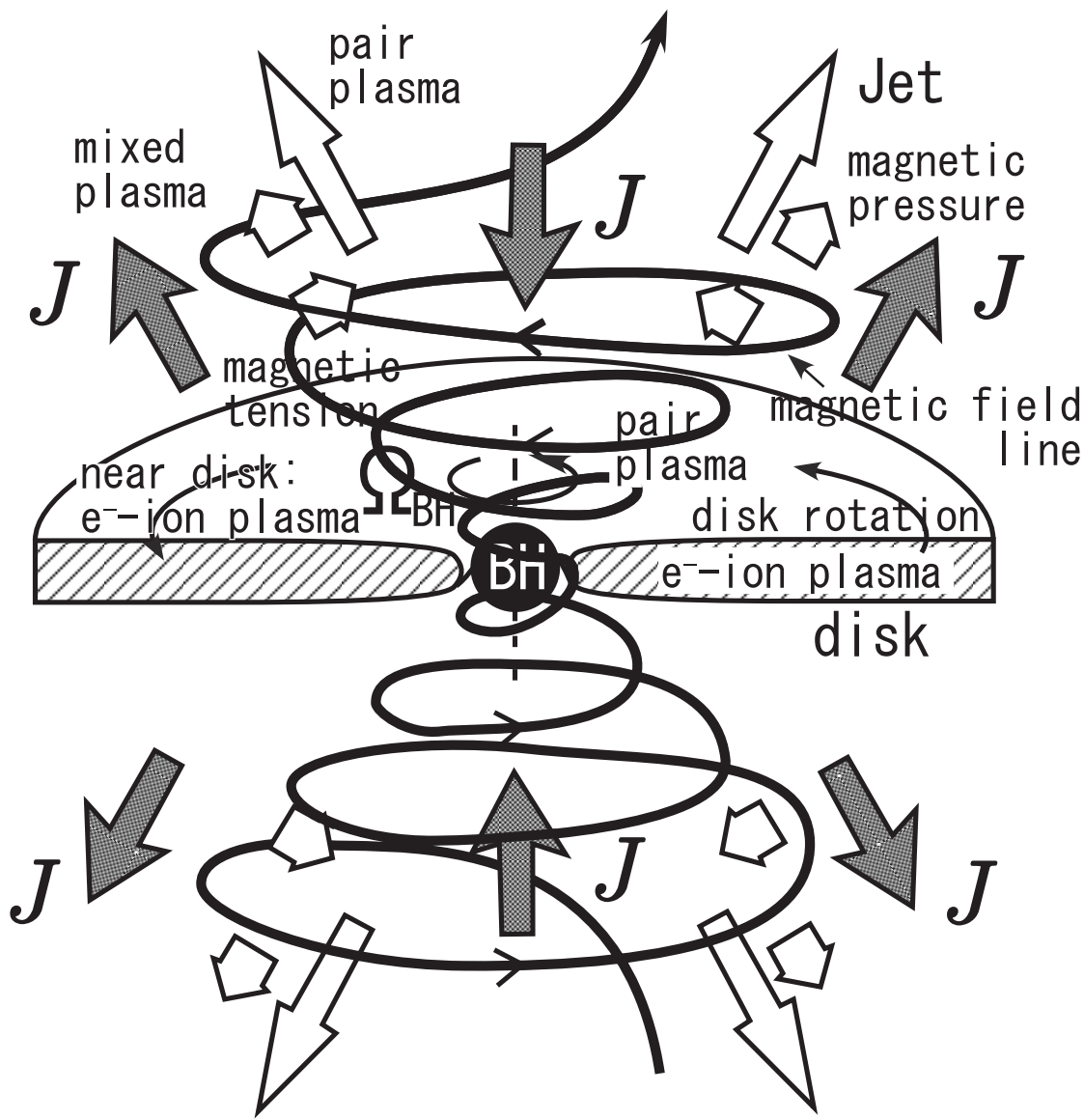}
\caption{Schematic picture of a relativistic jet forming region in a 
black hole magnetosphere of an AGN.
The disk is thought to be mainly composed of a electron-ion plasma, 
while the corona 
near the black hole and the core of the jet would be mainly of pair plasma.
The coronae near the disk and outer shell of the jet probably consist
of mixed plasma with ions, electrons, and positrons.
Magnetic field lines are twisted by the accretion disk rotating
around the black hole, and the jet is driven by the magnetic pressure.
In this case, the current and return current are formed around the jet,
which is indicated by the arrows of the current density $\VEC{J}$.
\label{fig1}}
\end{figure}

We investigate whether unique properties of the electron-ion and pair plasmas shown by
their generalized RMHD equations are significant in a jet forming region 
around an AGN black hole, which has not been done by previous works with
standard RMHD equations. 
Here, we assume that in such a region global magnetic field crosses 
an accretion disk around a black hole and the plasma around the black hole
has zero resistivity, which is a typical
assumption for jet formation models (see Fig.\ref{fig1}) 
\cite{blandford82,shibata86,kudoh98,koide98,koide99,koide00}. 
The disk twists the magnetic field lines
rapidly near the black hole to increase the magnetic pressure.
The magnetic pressure (and tension) blows off the plasma near the disk.
The outflow of the plasma is collimated by the magnetic tension to
form a jet. We here make a rough estimation of the
distinct properties of the electron-ion and pair plasmas through the process 
around the black hole. 
For simplicity, we ignore the general and special relativistic
effects such as the frame-dragging effect of the rotating black hole
and the inertia effect of the magnetic field. 
The former effects are discussed in section \ref{sec8}.
In this section, we use the MKSA units.

As shown in Fig. \ref{fig1}, in such a situation, current is induced
along the rotation axis of the disk, and the return current is formed.
The generalized RMHD equations suggest that the current carries the momentum
to influence the jet formation. We evaluate the influence as follows.
We assume the velocity of the disk rotation is the Keplerian velocity
$v_{\rm K} = \sqrt{GM_{\rm BH}/r}$, where $G$ is the gravitational
constant, $M_{\rm BH}$ is the mass of the black hole, and $r$ is the
distance from the center of the black hole to the jet forming region.
The azimuthal component of the twisted magnetic field $B_\phi$ is
estimated by
\[
\frac{B_\phi^2}{2\mu_0 \rho} = \frac{1}{2} v_{\rm K}^2 ,
\]
where $\rho$ is the mass density of the coronal plasma around 
the black hole.
On the other hand, the Ampere's law yields $\mu_0 J \sim B_\phi/r$.
As shown in section \ref{sec6} (equation (\ref{ratio_pecu1})), 
we evaluate the influence of 
current momentum and the Hall effect by the ratio
$[\mu |\VEC{J}|/(ne)]/|\VEC{U}|$ and $\mu \rho_{\rm e}/(ne \gamma)$.
The former ratio is estimated as
\[
\frac{1}{V} \frac{J}{ne} \mu \sim \frac{c}{r} \sqrt{\frac{\epsilon_0 m}{ne^2}} \mu
= \frac{c\mu}{r \omega_{\rm p}}.
\]
In the pair plasma case, the Hall effect disappears. 
In the electron-ion plasma case, it is negligible when
\begin{equation}
\frac{c}{r \omega_{\rm p}} \Delta \mu \approx \frac{c}{r \omega_{\rm p}} \ll 1  .
\label{cond_hall}
\end{equation}
The critical particle number density, with which the current momentum 
or the Hall effect is significant, $n_{\rm crit}$, near the 
black hole is estimated by
\[
\frac{c}{r_{\rm S}} \sqrt{\frac{\epsilon_0 m}{n_{\rm crit} e^2}} 
\max(\mu, \Delta \mu) \sim 1  ,
\]
and using $\max(\mu, \Delta \mu) \leq 1$, we have
\[
n_{\rm crit} \sim \frac{c^2}{r_{\rm S}^2} \frac{\epsilon_0 m}{e^2} 
\{\max(\mu, \Delta \mu)\}^2
\leq \frac{c^6}{(2GM_{\rm BH})^2} \frac{\epsilon_0 m}{e^2}
=3.2 \times 10^{6} \left ( \frac{M_\sun}{M_{\rm BH}} \right )^2
\left [{\rm m}^{-3} \right ] .
\]
Here, we set $r=r_{\rm S}$ where $r_{\rm S}=2 G M_{\rm BH}/c^2$ is 
the Schwarzschild radius. 
In the case of an AGN, we calculate the critical number density
as $n_{\rm crit} \sim 3 \times 10^{-2} \left [ {\rm m}^{-3} \right ]$
when $M_{\rm BH} \sim 10^8 M_\sun$.
This critical number density $n_{\rm crit}$ is much less than
the actual value of the plasma around the black hole
and this evaluation shows insignificance of the current momentum 
and Hall effect.

We evaluate the inertia effect of the current in the Ohm's law (\ref{rmhdohm}).
As shown in section \ref{sec6} (equation (\ref{ratio_pecu2})), 
the influence is estimated by the ratio
\[
\frac{\mu}{c \tau} \frac{mn^2}{n^2e^2} \frac{J}{B} 
= \frac{\mu}{\tau \omega_{\rm p}} \frac{c}{r \omega_{\rm p}} 
\sim \left ( \frac{c}{r \omega_{\rm p}} \right )^2
\frac{v_{\rm K}}{c}  ,
\]
where $\tau$ is the characteristic time scale of the jet formation
$\tau \sim r/v_{\rm K}$.
This shows if the current momentum and the Hall effect are negligible,
we can neglect the current inertia effect because of equation 
(\ref{cond_hall}) and the condition $v_{\rm K} < c$.
As the results, the distinct properties of the pair and electron-ion plasmas
in the jet forming region near the black hole are not significant globally
in the AGN case.
Inversely, the condition that the current momentum is
significant is
\[
r \la  \frac{c}{\omega_{\rm p}} \mu.
\]
The Hall effect is not negligible when
\[
r \la \frac{c}{\omega_{\rm p}} \Delta \mu
\]
in the electron-ion plasma case, while it disappears in the pair plasma case.
The inertia of the current is significant in the case that
\[
\tau \la \mu \frac{1}{\omega_{\rm p}} 
\left ( \frac{c \omega_{\rm p}^{-1}}{r} \right ) .
\]
Here, we neglect the thermal electromotive force and the thermalized
energy exchange between the two fluids.
These conditions can be satisfied only in the small-scale phenomena.
The global plasmas of black hole magnetospheres 
may be influenced by these distinct effects through the
magnetic reconnection around the black holes \cite{koide08}.
This is discussed in section \ref{sec9} briefly.

\section{Generalized GRMHD equations for electron-ion/pair plasmas 
around black holes
\label{sec8}}

Around a foot-point of a jet near a black hole shown in Fig. 1,
the general relativistic effect becomes significant because of
strong gravity of the black hole.
In this section, we discuss the general relativistic effects of the
plasma dynamics near black holes, while in the previous sections
we investigated the RMHD equations in the special relativistic framework.
To treat the plasmas near black holes appropriately, 
we have to use the general relativistic equations. 
We get the equations by transforming
the generalized RMHD equations (\ref{rmhdnum})--(\ref{rmhdohm})
with the Maxwell equations
(\ref{4formfar}), (\ref{4formamp}) to
the covariant form of the general relativity
\begin{eqnarray}
&&\nabla_\nu (\rho U^\nu) = 0 , \label{grmhdnum} \\
&&\nabla_\nu  \left [ 
h \left (U^\mu U^\nu + \frac{\mu}{q^2} J^\mu J^\nu \right ) \right ]
= -\nabla^\mu p + J^\nu {F^\mu}_\nu ,  \label{grmhdmot} \\
&& \frac{1}{q}  \nabla_\nu \left [  
\frac{\mu h}{q} (U^\mu J^\nu + J^\mu U^\nu ) 
- \frac{2\mu \Delta \mu h}{q^2} J^\mu J^\nu \right ] \nonumber \\
&&= \frac{1}{2q} \nabla^\mu (\Delta \mu p - \Delta p) +
\left ( U^\nu - \frac{\Delta \mu}{q} J^\nu \right) {F^\mu}_\nu   
- \eta [J^\mu + Q (1+\Theta) U^\mu], \label{grmhdohm0} \\
&& \nabla_\nu \hspace{0.3em} ^*F^{\mu\nu} = 0 , \label{grmhdfar} \\ 
&& \nabla_\nu F^{\mu\nu} = J^\mu , \label{grmhdamp}
\end{eqnarray} 
where $\nabla_\mu$ is the covariant derivative \cite{misner70,weinberg72}.
Using the equation derived by the Maxwell equations
\[
(\nabla_\nu F_{\mu \sigma}) F^{\nu \sigma} = \frac{1}{4} g_{\mu\nu}
\nabla^\nu (F^{\kappa\lambda} F_{\kappa\lambda}) ,
\]
and equation (\ref{grmhdamp}), the equation of motion (\ref{grmhdmot})
becomes
\begin{equation}
\nabla_\nu T^{\mu\nu} = 0 ,
\end{equation}
where 
\[
T^{\mu\nu} = p g^{\mu\nu} 
+ h \left ( U^\mu U^\nu + \frac{\mu}{q^2} J^\mu J^\nu \right )
+ {F^\mu}_\sigma F^{\nu \sigma} - \frac{1}{4} g^{\mu\nu} 
(F^{\kappa\lambda}F_{\kappa\lambda}) .
\]
This equation corresponds to the equation of motion in the standard GRMHD,
for example, equation (A2) in Appendix A of Koide et al. (2006).
The newly additional term is only that of the current momentum
density $\mu J^\mu J^\nu/q^2$.
The Ohm's law becomes
\begin{eqnarray}
&&  \nabla_\nu \left [  
\frac{\mu h}{q^2} (U^\mu J^\nu + J^\mu U^\nu ) 
- \frac{2\mu \Delta \mu h}{q^3} J^\mu J^\nu \right ] \nonumber \\
&&= \frac{1}{2q} \nabla^\mu (\Delta \mu p - \Delta p) +
\left ( U^\nu - \frac{\Delta \mu}{q} J^\nu \right) {F^\mu}_\nu   
- \eta [J^\mu + Q (1+\Theta) U^\mu], \label{grmhdohm}
\end{eqnarray}
when we ignore the non-uniformity and time variation of $q$.
As shown in section \ref{sec3}, $H=(\eta q)^2/(\mu h)$ should be
smaller than several (almost unity) because of causality. 
Equation (\ref{grmhdohm})
with $H \la 1$ is called the equation of the ``generalized general-relativistic 
Ohm's law" and equations (\ref{grmhdnum}), (\ref{grmhdmot}), (\ref{grmhdohm}),
(\ref{grmhdfar}), (\ref{grmhdamp})
with $H \la 1$ are called the ``generalized GRMHD equations".
Equation (\ref{grmhdohm}) is drastically different from the ideal MHD condition
$U^\nu {F^\mu}_\nu = 0$, which is used customarily in present GRMHD 
simulations. To show the plasma dynamics governed by the generalized 
GRMHD equations, we discuss them using the 3+1 formalism shown by
Koide, Kudoh, \& Shibata (2006) (see Appendix A of the paper;
definition of variables therein).
The variable observed by the ``zero angular momentum observer (ZAMO)" 
frame is denoted by a hat.
The energy density observed by the ZAMO frame is
\[
\epsilon + \gamma \rho = \hat{T}^{00} 
= h \gamma^2 + \frac{\mu}{q^2} \rho_{\rm e}^2 - p
+ \frac{\hat{B}^2}{2} + \frac{\hat{E}^2}{2}  ,
\]
where $\epsilon + \gamma \rho$ is the total energy density.
This shows ``electric charge" has the rest mass energy,
which is disregarded in the standard GRMHD.
The three components of the momentum density,
\[
\hat{P}^i = \hat{T}^{i0} =
h \left ( \gamma \hat{U}^i + \frac{\mu}{q^2} \hat{\rho}_e \hat{J}^i \right )
+ (\hat{\VEC{E}} \times \hat{\VEC{B}})_i ,
\]
show that the current has the momentum.
The stress-tensor,
\[
\hat{T}^{ij} = p \delta^{ij} + h \left (
\hat{U}^i \hat{U}^j + \frac{\mu}{q^2} \hat{J}^i \hat{J}^j \right )
+ \left ( \frac{\hat{B}^2}{2} + \frac{\hat{E}^2}{2} \right ) \delta^{ij}
-\hat{B}_i \hat{B}_j - \hat{E}_i \hat{E}_j  ,
\]
indicates the current also carries the momentum.
Summarizing the above results, we find that 
the difference of equation of motion between the generalized GRMHD and 
standard GRMHD equations is only with respect to the inertia and
momentum of the current, which are found in the special relativistic MHD
equations.

We next examine the Ohm's law. Defining
\[
T_{\rm ohm}^{\mu\nu} =
\frac{\mu h}{q^2} (U^\mu J^\nu + J^\mu U^\nu)
- \frac{2 \mu \Delta \mu h}{q^3} J^\mu J^\nu ,
\]
we have another formalism of the general relativistic generalized Ohm's law
\begin{equation}
\nabla_\nu T_{\rm ohm}^{\mu\nu} = \frac{1}{2q} \nabla^\mu (\Delta \mu p - \Delta p)
+ \left ( U^\nu - \frac{\Delta \mu}{q} J^\nu \right ) {F^\nu}_\mu
-\eta [J^\mu + Q(1+\Theta) U^\mu]  .
\label{genrelgenohm4}
\end{equation}
The $i$-th component of equation (\ref{genrelgenohm4}) yields the 3+1 form
\[
\frac{\partial}{\partial t} \left ( \frac{\mu h}{q^2} \hat{J}^{\dagger j} \right ) = 
-\frac{1}{h_1 h_2 h_3} \sum_j  \partial_j \left [ 
\frac{\alpha h_1 h_2 h_3}{h_j} (\hat{T}_{\rm ohm}^{ij} 
+  \beta^j \frac{\mu h}{q^2} \hat{J}^{\dagger i} ) \right ]
- \frac{\mu h}{q^2} \frac{1}{h_i} 
\frac{\partial \alpha}{\partial x^i} \rho_{\rm e}^\dagger
\]
\begin{equation}
+ \alpha f_{\rm ohm}^i - \sum_j \frac{\mu h}{q^2} \sigma_{ji} \hat{J}^{\dagger j}
+ \sum_j \alpha \beta^j \frac{\mu h}{q^2} \left (G_{ij} \hat{J}^{\dagger i}
- G_{ji} \hat{J}^{\dagger j} \right )
+\frac{1}{2q} \frac{1}{h_i} \frac{\partial}{\partial x^i}
(\Delta \mu p - \Delta p) 
\label{genrelgenohm3+1}
\end{equation}
\[
+ \gamma [E^i + (\VEC{V} \times \VEC{B})_i]
-\frac{\Delta \mu}{q} \left [ \rho_{\rm e} E^i + (\VEC{J} \times \VEC{B})_i \right ]
-\eta [J^i + Q(1+\Theta) U^i],
\]
where $h_1^2$, $h_2^2$, $h_3^2$ are spacial diagonal elements of the metrics, 
$\beta^1$, $\beta^2$, $\beta^3$ are the shift vectors, $\alpha$ is 
the lapse function,
$\sigma_{ij}$ is the shear of the frame-dragging 
$\sigma_{ij}=(1/h_j)[\partial (\alpha \beta^i)/\partial x^j]$,
$f_{\rm ohm}^i = \sum_j (G_{ij} \hat{T}_{\rm ohm}^{ij} 
- G_{ji} \hat{T}_{\rm ohm}^{jj})$, and 
$G_{ij}=-1/(h_i h_j) (\partial h_i/\partial x^j)$.
Here, $\rho_{\rm e}^\dagger \equiv (q^2/\mu h) \hat{T}_{\rm ohm}^{00} = 
2\rho_{\rm e}(\gamma - \Delta \mu \rho_{\rm e}/(en))$ and 
$\hat{J}^{\dagger i} \equiv (q^2/\mu h) \hat{T}_{\rm ohm}^{0i} 
= \gamma \hat{J}^i + \rho_{\rm e} \hat{U}^i -(2\Delta \mu/q) \rho_{\rm e} \hat{J}^i$ 
are on the order of the charge density and the current density, respectively,
and $\alpha$ is regarded as gravitational potential.
With respect to the term 
$-(\mu h /q^2)(1/h_i) (\partial \alpha/\partial x^i) \rho_{\rm e}^\dagger$, 
it clearly shows the gravity influences the Ohm's law
and behaves like electromotive force as Khanna (1998) pointed out.
Near the black hole where the gravitational acceleration is very strong,
such ``gravitational electromotive force" becomes significant when
the charge separation of the plasma occurs.
This effect is also found in the Newtonian two-fluid model
when we consider the gravitation term in equation of motion of the two fluids.
The forth and fifth terms of the right hand side of equation 
(\ref{genrelgenohm3+1}) 
\[
- \sum_j \frac{\mu h}{q^2} \sigma_{ji} \hat{J}^{\dagger j}
+ \sum_j \alpha \beta^j \frac{\mu h}{q^2} \left (G_{ij} \hat{J}^{\dagger i}
- G_{ji} \hat{J}^{\dagger j} \right )
\]
is also regarded as the effective electromotive force.
In the electromotive force, the shear of the frame dragging around 
black holes, $(\mu h/q^2) \sigma_{ji}$ and $(\mu h/q^2) \alpha \beta^j G_{ik}$,
corresponds to the electric resistivity.
This effective resistivity due to the frame-dragging shear is a pure
general relativistic effect, which may cause
the dynamo effect and magnetic reconnection near black holes
\cite{khanna94,khanna96,koide08}.

\section{Summary and discussion \label{sec9}}

In this paper, we derived the one-fluid equations of 
the two-component plasma from the relativistic two-fluid equations. 
On the basis of the new one-fluid equations and 
the dispersion relation analysis of the electromagnetic wave,
we proposed a set of generalized RMHD equations 
(\ref{rmhdnum})--(\ref{rmhdohm}) with Maxwell equations 
(\ref{4formfar}) and (\ref{4formamp}) which are applicable to
both pair and electron-ion plasmas. 
The generalized RMHD equations yielded the RMHD equations for the electron-ion
plasma (equations (\ref{normalrmhdnum})--(\ref{normalrmhdohm})) and the pair plasma
(equations (\ref{pairrmhdnum})--(\ref{pairrmhdohm})). 
The comparison between these sets of equations and the standard equations
(\ref{standrmhdnum})--(\ref{standrmhdohm})
clearly show the distinct properties of the electron-ion and pair plasmas.
We investigated the linear modes 
propagating in the pair and electron-ion plasmas using the generalized RMHD equations.
In the analysis, we found the effect of polarity rotation of Alfven wave
in the electron-ion plasma, while no rotation in the pair plasma.
This effect is similar to the Faraday rotation of the electromagnetic wave
propagating along magnetic field lines in a plasma, but not identical.
We also found other plausible unique properties of the electron-ion and pair plasmas,
such as the plasma oscillation, fast wave, and electromagnetic wave,
which also show the validity of the generalized RMHD equations.
With respect to the non-linear effects, we evaluated the distinct properties
of the pair and electron-ion plasmas with the significance of each distinct term
of the generalized RMHD equations and with the special case of
black hole magnetospheres
where jets are formed. It confirmed the validity of the standard 
RMHD equations (\ref{standrmhdnum})--(\ref{standrmhdohm}), 
(\ref{4formfar}), (\ref{4formamp}) 
on the global phenomena in black hole magnetospheres of AGNs. 
We also discussed the generalized GRMHD equations (\ref{grmhdnum})--(\ref{grmhdamp}) 
and revealed the unique effects of the plasmas around black holes.
As indicated by Khanna (1998),
we confirmed the appearance of the ``gravitational electromotive force"
due to the strong gravity of a black hole when the plasma has charge separation.
We newly found the effective resistivity due to shear of frame dragging
around a rotating black hole.
To investigate the non-linear phenomena qualitatively, numerical
method is required and should be developed in the near future.

According to the results in subsection \ref{sec7bdc}, 
if the relative velocity of the two fluids and the thermal energy
of the fluids are non-relativistic and their proper particle number densities
are approximately equal, $\gamma' \approx 1$, $p_\pm \ll \rho_\pm$, $n_+ \approx n_-$, 
the generalized RMHD equations are applicable.
Otherwise, the premise condition ($\Delta h^\dagger \ll h^\dagger$) 
of the generalized RMHD equations require $p_+/\rho_+ \approx p_-/\rho_-$,
that is, $T_+ = (m_+/m_-) T_-$. 
This condition is implausible for the electron-ion plasma
when the ion temperature is not much larger than
the electron temperature. 
In the pair plasma case, it yields $p_+/n_+ = T_+ \approx T_-=p_-/n_-$.
The special condition of the electron-ion plasma, $T_+ = (m_+/m_-) T_-$ may be
possible for optically thin Advection Dominated Accretion Flows
(ADAFs) in the electron-ion plasma of the accretion disk around black holes
\cite[e.g.,][see Table 10.1]{kato98}.
If these premise conditions are not satisfied, the generalized RMHD
equations, which are reducible to the standard RMHD equations,
are not applicable to the plasma. That is, a relativistically hot plasma
($p_\pm \ga \rho_\pm$) can not be treated by RMHD
if $p_+ \approx p_- (m_+/m_-)$ and $n_+ \approx n_-$ are not satisfied.
In such a case, the generalized RMHD approximation breaks down.
Then, the generalized RMHD equations 
should be replaced by more appropriate equations
for the plasma with relativistic internal energy
(relativistic high temperature, relativistically fast relative velocity
of the two fluids).
The standard RMHD equations, which have been utilized indiscriminately 
in astrophysics, are applicable to the astrophysical objects when the premise
of the generalized RMHD equations are satisfied and the unique properties of 
the electron-ion/pair plasma are negligible.
These conditions indicated in this paper depend on 
the averaging procedure of the variables (\ref{aveenth1}). 
The averaging of the variables (\ref{aveenth1}) enables us to derive 
the one-fluid equations (\ref{onefluidnum})--(\ref{onefluidohm})
from the two-fluid equations 
(\ref{4formnum})--(\ref{4formmom}) relatively easily.
However, we have to check the validity of the averaging method.
To confirm it, the Vlasov--Boltzmann equation of the plasma should be utilized.
This task is beyond the scope of this paper, while it is interesting and
important. It should be investigated in the near future.

As shown in subsection \ref{sec7bhj}, distinct properties 
of the electron-ion and pair plasmas
do not influence the global dynamics of the plasmas in black hole
magnetospheres of AGNs because they are large-scale RMHD phenomena. 
On the other hand, local phenomena such as magnetic reconnection may 
change the influence to the global dynamics drastically,
because they are possible to change the magnetic configuration globally, although
the reconnection region is much smaller than the global scale of 
magnetospheres. In such a small region, the distinct properties of the electron-ion
and pair plasmas may become significant. For example, the inertia effect
of the current plays a role to cause the magnetic diffusion
instead of the electric resistivity.
This distinct magnetic reconnection may be important in the pair plasma,
which is thought to locate at the coronal regions near black holes 
of AGNs and GRBs.

With respect to the corona between a black hole and its disk, 
in generally speaking,
the plasmas would be composed of electrons, ions, and positrons,
while the amount of positrons in the plasmas around black holes has not 
been confirmed observationally yet.
This is regarded as a mixture of electron-ion plasma and pair plasma.
The simplest treatment of the mixed plasma is given by
$ m_- = m_{\rm e}$ and $ m_+ = m_{\rm e} + (m_{\rm i} - m_{\rm e}) \xi$,
where $m_{\rm i}$ is the mass of the ion and $\xi = n_{\rm i}/n_{\rm e^-}$
is the relative proportion of ions ($n_{\rm i}$ and $n_{\rm e^-}$ 
are the particle number densities of ions and electrons, respectively).
Here, we assumed the local (proper) charge neutrality, 
$n_{\rm e^-} = n_{\rm i} + n_{\rm e^+}$, where $n_{\rm e^+}$
is the number density of the positrons.
In this treatment, the relative proportion of ions $\xi$ should be
given by other condition. For example, we may be able to trace 
the plasma element where $\xi$ is constant. However, when the relative
velocity between the positron fluid and the ion fluid is significant,
$\xi$ on a certain plasma element changes. In such a case, the treatment
becomes difficult.
The advanced formalism of the mixed plasma is a forthcoming
and important subject in the relativistic plasma physics.
Additionally, other effects such as pair creation/annihilation, radiation, 
atomic processes, and more general EoS should be considered.
Furthermore, as we mentioned above, for the relativistically hot plasma
and relativistically strong current plasma, the premise of the generalized
RMHD equations breaks down.
In plasmas at black hole magnetospheres, it is not surprising that
the temperature and current density are relativistically high and strong.
To deal with such relativistic plasmas, we have to develop a new
scheme beyond the present generalized RMHD equations.


\acknowledgments

I am grateful to Mika Koide, David L. Meier, Masaaki Takahashi, 
and Takahiro Kudoh for their helpful comments on this paper.
This work was supported in part by the Science Research Fund of
the Japanese Ministry of Education, Culture, Sports, Science, and Technology.

\appendix

\section{Frictional four-force density of two fluids
\label{appenda}}

We assume a relative velocity of two fluids is 
non-relativistic so that $M=U_\nu U^\nu = -1$.
Then, when we observe the fluids from the local center-of-mass
frame S', the fluids can be treated as non-relativistic fluids.
The four-force density of the friction from the negatively charged
fluid to the positively charged fluids is
\begin{equation} 
{f_+^{i}}' = -m_+\sigma_{+-} v_{\rm r} n_+ n_- \gamma_+^\prime \gamma_-^\prime 
(V_+^{i \prime} -V_-^{i \prime}), 
\end{equation}
where $\sigma_{+-}$ is the cross-section of the collision between
the positive particle and negative particle, $v_{\rm r}$ is the average 
relative velocity between the two fluids including the thermal velocity.
Similarly, the frictional four-force density of the positively charged 
fluid to the negatively charged fluid is
\begin{equation} 
{f_-^{i}}' = -m_-\sigma_{-+} v_{\rm r} n_+ n_- \gamma_+^\prime \gamma_-^\prime 
(V_-^{i \prime} -V_+^{i \prime}). 
\end{equation}
According to action-reaction principle, we have 
\begin{equation}
m_+ \sigma_{+-} = m_- \sigma_{-+}.
\end{equation}
Then we define the following variables,
\begin{equation}
A = \eta e^2 = m_+ \sigma_{+-} v_{\rm r} = m_- \sigma_{-+} v_{\rm r}.
\end{equation}
In the center-of-mass frame S', we have
\begin{equation}
m_+ n_+ {U_+^i}' + m_- n_- {U_-^i}' = 0.
\end{equation}
When we consider the Lorentz transformation $x^\mu = {b^\mu}_\nu {x^\nu}'$,
we find
\begin{equation}
{b^\mu}_0 = \frac{U^\mu}{\gamma'}.
\end{equation}
We calculate the frictional four-force density by
\[
f_+^\mu = {b^\mu}_\nu {f_+^\nu}' = \frac{U^\mu}{\gamma'}{f_+^0}'
+ {b^\mu}_i {f_+^i}' = \frac{U^\mu}{\gamma'} {f_+^0}' 
-\frac{A \rho}{me} \sqrt{-M} \left [ 
J^\mu - \frac{Q}{M} U^\mu \right ] ,
\]
where $M=U_\nu U^\nu = - \gamma'^2$ and $Q=U_\nu J^\nu = - \gamma' {J^0}'$.
We also note that
\[
{f_+^i}' {v_{+i}}' = - \frac{A \rho m_-}{m^2 n_+ \gamma_+' \gamma' e^2}
(Q^2 - MW) < 0   ,
\]
where $W=J_\nu J^\nu$ and
\[
{f_-^i}' {v_{-i}}' = - \frac{A \rho m_-}{m^2 n_- \gamma_-' \gamma' e^2}
(Q^2 - MW) .
\]
We consider the thermal energy gain rate of the positive charged fluid,
\[
{f_+^0}' = 
\frac{-{f_+^i}' {v_{+i}}' \theta_+ - {f_-^i}' {v_{-i}}' \theta_-}
{\gamma_+' n_+ + \gamma_-' n_-} n_+ \gamma_+'
-{f_+^i}' {v_{+i}}' (1 - \theta_+) - (-{f_+^i}' {v_{+i}}')
\]
\[
=\frac{-1}{\gamma_+' n_+ + \gamma_-' n_-}
(\theta_- n_+ \gamma_+' {f_-^i}' {v_{-i}}' - 
\theta_+ n_- \gamma_-' {f_+^i}' {v_{+i}}' )
\]
\[
=\frac{A\rho}{me^2} 
\frac{\theta_+ m_- [\rho M - (m_+/e)Q ]^2 - \theta_- m_+ [\rho M + (m_-/e)Q ]^2}
{\left ( 2\rho M + \frac{m_- - m_+}{e}Q \right )
\left ( \rho M - \frac{m_+}{e}Q \right )
\left ( \rho M + \frac{m_-}{e}Q \right )} (Q^2 -MW),
\]
where $\theta_+$, $\theta_-$ are the redistribution 
coefficient of the thermalized energy to the positively charged fluid
and negative one with the equipartition principle, respectively.
When $\theta = \theta_+ = \theta_-$, we obtain
\begin{equation}
{f_+^0}' = - \frac{A \rho}{me} Q \Theta \le 0 ,
\end{equation}
where 
\begin{equation}
\Theta = - \frac{\theta}{eQ}
\frac{(Q^2-MW) \left [ (m_+ - m_-) \left \{ 
-(\rho M)^2 + \left ( \frac{Q}{e} \right )^2 m_+ m_- \right \}
-\frac{4\rho M Q m_+ m_-}{e} \right ]}
{\left ( 2\rho M + \frac{m_- - m_+}{e}Q \right )
\left ( \rho M - \frac{m_+}{e}Q \right )
\left ( \rho M + \frac{m_-}{e}Q \right )} .
\end{equation}
Using $\Theta$, we finally obtain the frictional four-force density,
\begin{equation}
f_+^\mu = - \frac{A \rho}{me} \sqrt{-M} \left [ 
J^\mu - \frac{Q}{M} (1+ \Theta) U^\mu \right ] .
\end{equation}
Because $R^\mu = f_+^\mu$, we obtain the Ohm's law,
\begin{eqnarray}
&& \frac{1}{2ne}   \partial_\nu  \left [  
\frac{h^\ddagger}{2ne} (U^\mu J^\nu + J^\mu U^\nu ) + \Delta h^\dagger U^\mu U^\nu
+ \frac{\Delta h^\sharp}{(2ne)^2} J^\mu J ^\nu \right ] \\
&=& \frac{1}{2ne} \partial^\mu (\Delta \mu - \Delta p) +
\left ( U^\nu - \frac{\Delta \mu}{ne} J^\nu \right) {F^\mu}_\nu  
- \eta \sqrt{-M} \left [ J^\mu - \frac{Q}{M} (1+\Theta) U^\mu \right ] .
\end{eqnarray}
Here, we recognize $\eta$ is the resistivity.


\end{document}